\documentclass[11pt,preprint]{aastex}

\newcommand{\be}{\begin{equation}}
\newcommand{\ee}{\end{equation}}
\newcommand{\bea}{\begin{eqnarray}}
\newcommand{\eea}{\end{eqnarray}}

\def\cm   {\,{\rm cm}}
\def\tauLar {\tau_{\rm Lar}}     % Larmor time
\def\Pgas  {P_{\rm gas}}
\def\Pmag  {P_{\rm mag}}
\def\mH    {m_{\rm H}}
\def\nH    {n_{\rm H}}
\def\s     {\,{\rm s}}
\def\Vfive {V_5}     
\def\Lten  {L_{10}}     
\shortauthors{Yan, Lazarian \& Draine}
\begin{document}

\title{Dust Dynamics in Compressible MHD Turbulence}

\author{Huirong Yan\altaffilmark{1}, A.  Lazarian\altaffilmark{1} and B. T. Draine\altaffilmark{2}}

\altaffiltext{1}{Department of Astronomy, University of Wisconsin, 475 N. Charter
St., Madison, WI 53706; yan, lazarian@astro.wisc.edu}

\altaffiltext{2}{Princeton University Observatory, Peyton Hall, Princeton, NJ 08544; draine@astro.princeton.edu}

\begin{abstract}
We calculate the relative grain-grain motions arising from interstellar
magnetohydrodynamic (MHD) turbulence. The MHD turbulence includes
both fluid motions and magnetic fluctuations. While the fluid motions
accelerate grains through hydro-drag, the electromagnetic fluctuations
accelerate grains through resonant interactions. We consider both
incompressive (Alfv\'{e}n) and compressive (fast and slow) MHD modes
and use descriptions of MHD turbulence obtained in Cho \& Lazarian
(2002). Calculations of grain relative motion are made for realistic grain charging and
interstellar
turbulence that is consistent with the velocity dispersions
observed in diffuse gas,
including
cutoff of the turbulence from various damping processes.
We show that fast modes dominate grain acceleration, and can drive
grains to supersonic velocities. Grains are also scattered by gyroresonance 
interactions, but the scattering is less important than acceleration
for grains moving with sub-Alfv\'{e}nic velocities. Since the grains are preferentially accelerated with large pitch
angles, the supersonic grains will be aligned with long axes perpendicular
to the magnetic field. We compare grain velocities arising
from MHD turbulence with those arising from photoelectric emission,
radiation pressure and H$_{2}$ thrust. We show that for typical interstellar
conditions turbulence should prevent these mechanisms from segregating
small and large grains. Finally, gyroresonant acceleration is bound to preaccelerate
	  grains that are further accelerated in shocks.  Grain-grain
	  collisions in the shock may then contribute to the
	  overabundance of refractory elements in the composition of
	  galactic cosmic rays. 
\end{abstract}


\section{Introduction}

Dust is an important constituent of the interstellar medium (ISM).
It interferes with observations in the optical range, but provides
an insight to star-formation activity through far-infrared radiation.
It also enables molecular hydrogen formation and traces the magnetic
field via emission and extinction polarization (see reviews by Hildebrand
et al. 2000, Lazarian 2000, 2002). The basic properties of dust (extinction, polarization, etc.) strongly depend on its size distribution. The latter
evolves as the result of grain collisions, whose frequency and consequences depend on grain
relative velocities (see discussions in Draine 1985, Lazarian \& Yan
2002ab).

Grain-grain collisions can have various outcomes, e.g., coagulation,
cratering, shattering, and vaporization. For collisions with $\delta v\leq10^{-3}$km/s,
grains are likely to stick or coagulate, as the potential energy due
to surface forces exceeds the initial center of mass kinetic energy.
Coagulation is considered the mechanism to produce large grains in
dark clouds and accretion disks. Collisions with $\delta v\geq20$km/s
have sufficient energy to vaporize at least the smaller of the colliding grains (Draine 1985). 
It is likely that some features of the grain
distribution, e.g., the cutoff at
large size (e.g., Kim, Martin \& Hendry 1994), are the result of fragmentation (Biermann \& Harwit 1980).
Even low-velocity grain collisions may have dramatic consequences
by triggering grain mantle explosion (Greenberg \& Yencha 1973, 
	  Schutte \& Greenberg 1991).

Various processes can affect the velocities of dust grains. Radiation,
ambipolar diffusion, and gravitational sedimentation all can bring
about a dispersion in grain velocities. It is widely believed that,
except in special circumstances (e.g., near a luminous young star,
or in a shock wave), none of these processes can provide substantial
random velocities so as to affect the interstellar grain population
via collisions (Draine 1985), except for possibly enhancing the coagulation rate.
Nevertheless, de Oliveira-Costa et al. (2002) speculated
that starlight radiation can produce
the segregation of different sized grains that was invoked to explain the imperfect
correlation of the microwave and $100\mu$m signals of the
foreground emission (Mukherjee et al. 2001). If true it has important implications
for the CMB foreground studies. However, the efficiency of this segregation
depends on grain random velocities, which we study in this paper.

The interstellar medium is magnetized and turbulent (see Arons \&
Max 1975, Scalo 1987, Lazarian 1999). Although turbulence has been
invoked by a number of authors (see Kusaka et al. 1970, V\"olk
et al. 1980, Draine 1985, Ossenkopf 1993, Weidenschilling \& Ruzmaikina
1994) to provide substantial grain relative motions, the turbulence
they discussed was not magnetized. Dust grains are charged, and their
interactions with magnetized turbulence are
very different from the
hydrodynamic case. Lazarian \& Yan (2002a, henceforth LY02 and
2002b) applied the theory of Alfv\'{e}nic turbulence (Goldreich
\& Sridhar 1995, henceforth GS95, see Cho, Lazarian \& Vishniac
2002a for a review) to grain acceleration and considered the motions
that emerge due to incomplete coupling of grains and gas. Unlike the
pure hydrodynamic case discussed by earlier authors, LY02 took into
account that the motions of grains are restricted by magnetic fields
in the direction perpendicular to the field lines, and also took into
account the anisotropy of Alfv\'{e}nic turbulence.

While Alfv\'{e}nic turbulence is the turbulence in an incompressible fluid, the
ISM is highly compressible. Compressible MHD turbulence has been studied
recently (see review by Cho \& Lazarian 2003a and references therein).
In Yan \& Lazarian (2003, henceforth YL03) we identified a new mechanism
of grain acceleration-- gyroresonance --that is based on the
direct interaction of charged grains with MHD turbulence. YL03 provided a
test calculation of grain acceleration in compressible MHD turbulence, both by hydro-drag and by gyroresonance (see also review by Lazarian \& Yan 2003).

In what follows, we describe grain acceleration by MHD turbulence
in different phases of the ISM. Solving the Fokker-Planck equation
including simultaneously the hydro-drag and gyroresonance would be
a formidable task that we do not attempt here. Instead, we try to
simplify the problem by separating these two processes. For the random
fluid drag, we use a simple scaling 
argument similar to the
approach in Draine (1985) and LY02. While dealing with gyroresonance, 
we follow the approach adopted in YL03, i.e.,
we do not include the motion of ambient gas and assume that turbulence
provides nothing but electromagnetic fluctuations. These approximations
should yield correct answers when one of the mechanisms is dominant.
When the accelerations arising from the two mechanisms are comparable
the situation is more complicated, as the gaseous friction that we
use for gyroresonance calculations will be, in general, affected by
fluid motions. We do not develop explicit theory for this case. But
taking into account that it is the motions at the decoupling scale
that accelerate grains via hydro-drag, we think it is reasonable to
estimate the velocity gains from the simultaneous action of hydro-drag and gyroresonance by adding them in quadrature. 

To describe the turbulence we use the statistics of Alfv\'{e}nic
modes obtained in Cho, Lazarian \& Vishniac (2002b, hereafter CLV02)
and compressive modes obtained in Cho \& Lazarian (2002, hereafter
CL02, 2003bc)\footnote{The limitations on the applicability of 
such an approach are described in Yan \& Lazarian (2003)}. 
We apply our results to different phases of ISM, including
the cold neutral medium (CNM), warm neutral medium (WNM), warm ionized
medium (WIM), molecular cloud (MC) and dark cloud (DC) conditions, to estimate the implications for
coagulation, shattering and segregation of grains.

In what follows, we introduce the statistical description of MHD turbulence
and damping processes ($\S2$), describe motions arising from hydro-drag
($\S3$) and gyroresonance ($\S4$), apply our results to various
ISM phases ($\S5$), discuss astrophysical implications of our results
($\S6$), and provide the summary in $\S7$.

\section{MHD cascade and its damping}

MHD perturbations can be decomposed into Alfv\'{e}nic, slow and fast
modes (see Alfv\'{e}n \& F\"{a}lthmmar 1963). Alfv\'{e}nic turbulence
is considered by many authors as the default model of interstellar
magnetic turbulence. This is partially motivated by the fact that
unlike compressive modes, the Alfv\'{e}nic modes are essentially free
of damping in a fully ionized medium (see Ginzburg 1961, Kulsrud \&
Pearce 1969).  Important questions arise. Can the MHD perturbations that characterize turbulence be separated into distinct modes? Can the linear modes be used for this purpose? The separation into Alfv\'en and pseudo-Alfv\'en modes is the cornerstone of the Goldreich-Sridhar (1995, henceforth GS95) model of turbulence. This model and the legitimacy of the separation were tested successfully with numerical simulations (Cho \& Vishniac 2000; Maron \& Goldreich 2001; CLV02). Separation of MHD perturbations in compressible media into fast, slow and Alfv\'en modes is discussed in GS95, Lithwick \& Goldreich 2001, and CL02. The actual decomposition of MHD turbulence into Alfv\'en, slow and fast modes was performed in CL02, and Cho \& Lazarian
(2003, henceforth CL03), who also quantified the intensity of the interaction between different modes (see below).

Unlike hydrodynamic turbulence, Alfv\'{e}nic turbulence is anisotropic, with eddies elongated
along the magnetic field (see Montgomery \& Turner 1981, Shebalin, Matthaeus, \& Montgomery 1983, Higdon 1984, 
Zank \& Matthaeus 1992). This
happens because it is easier to mix the magnetic field lines perpendicular
to the direction of the magnetic field rather than to bend them. The
GS95 model describes \textit{incompressible} Alfv\'{e}nic turbulence,
which formally means that plasma $\beta\equiv \Pgas/\Pmag$, i.e., the
ratio of gas pressure to magnetic pressure, is infinity. The turbulent velocity spectrum is easily obtained. Calculations in CLV02 prove that motions
perpendicular to magnetic field lines are essentially hydrodynamic.
As the result, energy transfer rate due to those motions is a constant
$\dot{E_{k}}\sim v_{k}^{2}/\tau_{k}$, where $\tau_{k}$ is the energy
eddy turnover time $\sim(v_{k}k_{\perp})^{-1}$, where $k_{\perp}$is
the perpendicular component of the wave vector $\mathbf{k}$. The
mixing motions couple to the wave-like motions parallel to magnetic
field giving a critical balance condition, i.e., $k_{\bot}v_{k}\sim k_{\parallel}V_{A}$,
where $k_{\parallel}$ is the parallel component of the wave vector
$\mathbf{k}$, and $V_{A}$ is the Alfv\'{e}n speed. From these arguments,
the scale dependent anisotropy $k_{\parallel}\propto k_{\perp}$and
a Kolmogorov-like spectrum for the perpendicular motions $v_{k}\propto k^{-1/3}$
can be obtained. 

It was conjectured in Lithwick \& Goldreich (2001) that the GS95 scaling
should be approximately true for Alfv\'{e}n and slow modes in moderately
compressible plasma. For magnetically dominated (i.e., low $\beta$) 
plasma, CL02 showed that the coupling of Alfv\'{e}nic and
compressive modes is weak and that the Alfv\'{e}nic and slow modes
follow the GS95 spectrum. This is consistent with the analysis of
HI velocity statistics (Lazarian \& Pogosyan 2000, Stanimirovic \&
Lazarian 2001) as well as with electron density statistics (see
Armstrong, Rickett \& Spangler 1995). Calculations in Cho \& Lazarian
(2003, hereafter CL03) demonstrated that fast modes are marginally
affected by Alfv\'{e}n modes and follow acoustic cascade in both
high and low $\beta$ media. 

In what follows, we consider both Alfv\'{e}n
modes and compressive modes and use the description of those modes
obtained in CL02 and CL03 to study dust acceleration by MHD turbulence.

The distribution of energy between compressive and incompressive modes depends, in general, on the driving of turbulence. CL02 and CL03 studied generation of compressive perturbations using random incompressive driving,
obtaining an expression that relates the energy in fast $\sim\delta V_f^2$ and Alfv\'en $\sim\delta V_A^2$ modes, 

$$
(\delta V_f/\delta V_A)^2\sim \delta V_AV_A/(V_A^2+C_S^2),
$$
where $C_S$ is the sound speed. This relation testifies that at large scales incompressive driving can transfer an appreciable part of energy into fast modes. However, at smaller scales the drain of energy from Alfv\'en to fast modes is marginal. 
Therefore the cascades evolve without much cross talk. 
A more systematic study of different types of driving is required. In what follows we assume that equal amounts of energy are transfered into fast and Alfv\'en modes when driving is at large scales.

We show that while simple scaling relations are sufficient for obtaining
the velocities arising from hydro-drag, much more sophisticated tools
are necessary for calculating gyroresonance (see Yan \& Lazarian 2002, henceforth YL02, YL03). The corresponding statistics of turbulence is presented in Appendix B. 

At small scales the turbulence spectrum is truncated by damping. Various
processes can damp the MHD motions. In partially ionized plasma, the
ion-neutral collisions are the dominant damping process. In fully
ionized plasma, there are basically two kinds of damping: collisional
or collisionless damping (see Appendix A for details). Their relative
importance depends on the mean free path $l=v_{th}\tau=6\times10^{11}(T/8000K)^{2}\cm^{-2}/n$ in the ISM (Braginskii 1965). If the wavelength is larger than the mean
free path, viscous damping dominates. If, on the other hand, the wavelength
is smaller than mean free path, then the plasma is in the collisionless
regime and collisionless damping is dominant. 

To obtain the truncation scale, the damping time should be compared
to the cascading time. As we mentioned earlier, the Alfv\'{e}nic
turbulence cascades over one eddy turn over time $(k_{\perp}v_{k})^{-1}\sim(k_{\parallel}V_{A})^{-1}$.
The cascade of the fast modes is a bit slower: 
\begin{equation}\tau_{k}=\omega/k^{2}v_{k}^{2}=(k/L)^{-1/2}\times V_{f}/V^{2},\label{cascade}
\end{equation}
where $V_{f}$ is the phase velocity of fast waves, and $V$ is the turbulence
velocity at the injection scale (CL02). If the damping is faster than
the cascade, the turbulence is truncated. Otherwise, for the sake
of simplicity, we ignore the damping and assume that the turbulence
cascade is unaffected. According to CL02 the transfer of energy between
Alfv\'{e}n, slow and fast modes of MHD turbulence is suppressed. This
allows us to consider different components of MHD cascade independently.

\section{Grain Charge}

The net electrical charge on a grain 
is the result of competition between collisions with electrons, which
add negative charge, and photoelectric emission and collisions with
ions, which remove negative charge.
We assume the grains to be spheres consisting of either ``astronomical
silicate'' or graphite, with absorption cross sections calculated as
described by Weingartner \& Draine (2001a; henceforth WD01a), 
and photoelectric yields
(as a function of $Z$) estimated by 
Weingartner \& Draine (2001b; henceforth WD01b).

The grain charge depends on the electron density $n_e$.
While many previous studies have assumed cosmic ray ionization rates
$\zeta\approx 1\times10^{-17}{\rm s}^{-1}$ (e.g., Ruffle et al. 1998),
recent observational determinations (Black \& van Dishoeck 1991;
Lepp 1992; 
McCall et al.\ 2003) suggest 
$\zeta\approx 1-10\times10^{-16}{\rm s}^{-1}$ in H~I clouds.
We adopt an electron
density $n_e\approx 0.03\cm^{-3}$ for CNM conditions, consistent with
a detailed study of the ionization toward 23 Ori (Welty et al. 1999;
Weingartner \& Draine 2002), corresponding to
an H ionization rate $\zeta\approx1.5\times10^{-16}\s^{-1}$.

For the outer regions of molecular clouds (MC) we take $\nH\approx300\cm^{-3}$
and $n_e/\nH\approx10^{-4}$, mainly due to photoionization of metals,
where $G_{UV}=0.1$ is the UV intensity relative
to the average interstellar radiation field.
For ``dark clouds'' with $\nH\approx10^4\cm^{-3}$, we consider 
$G_{UV}/n_e=1\cm^3$, resulting in negatively-charged grains.
%---

Fig.~1a shows the mean grain 
charge $\langle Z\rangle$ for graphite and silicate grains
in various phases of the ISM (see Table 1).

The charge on a given grain fluctuates. 
Let $f_Z$ be the probability of the grain being in charge
state $Z$, and let $r_{Z}$ be the probability per unit time
of leaving charge state $Z$.
The characteristic time scale for the grain charge to fluctuate is 
$t_Z \equiv \langle (Z-\langle Z\rangle)^2\rangle 
/\sum_Z f_Z r_Z$. Fig.~1b compares
the charge fluctuation time
$t_Z$ to the Larmor time (= Larmor period/$2\pi$), 
$\tauLar\equiv1/\Omega=  m_{gr}c/\langle Z\rangle eB$ and the gas drag timescale $t_{\rm drag}^0$ for subsonic motion. 
We see that, except for $a\lesssim 10^{-6}\cm$ grains in dark clouds, the grain charge fluctuation time is much shorter than either of these dynamical times, so that these fluctuations can be ignored and the charge on a given grain can be assumed to be constant, equal to the time-averaged charge $\langle Z\rangle$.

\begin{figure*}
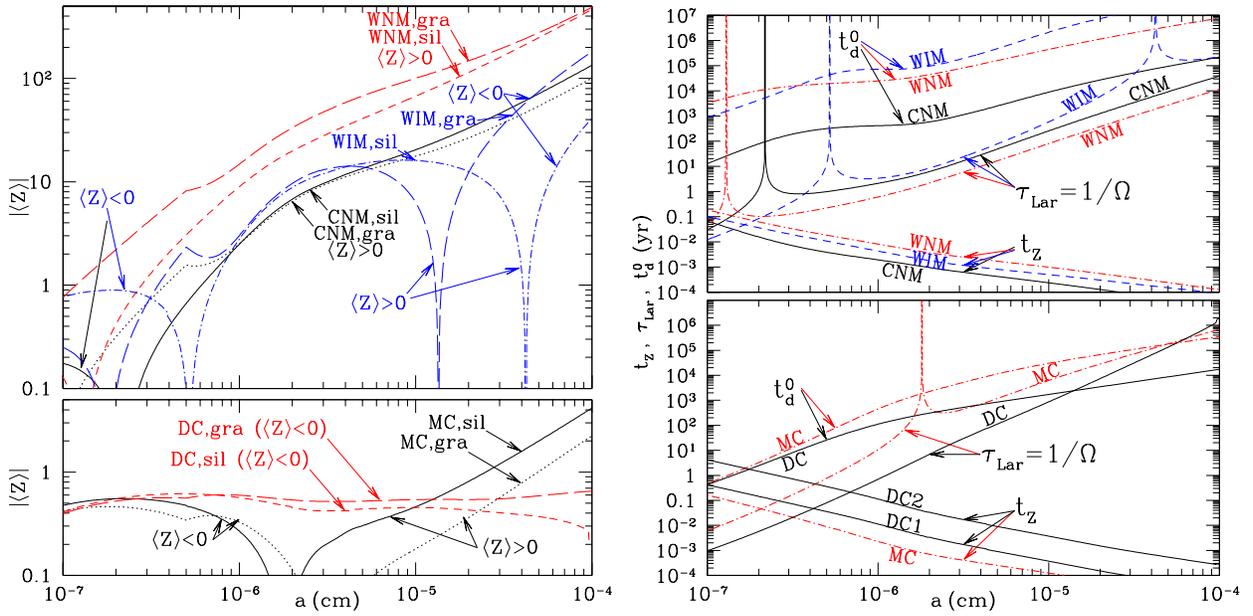

\includegraphics[width=0.49\textwidth]{f1a.cps} \hfil
\includegraphics[width=0.49\textwidth]{f1b.cps} 
\caption{\emph{left}: $|\langle Z\rangle|$ as a function of grain radius for carbon and silicate
grains in 6 different environments:  CNM,WNM,WIM,MC, DC1 and DC2. $|\langle Z\rangle|$ distribution is the same for DC1 and DC2, which are referred to as DC in the plot. 
\emph{right}: the gas-drag time $t_d^0$ for subsonic grains, the 
Larmor time $\tauLar$, and the charge relaxation time scale, 
$t_Z$, all as a function of
grain size for silicate grains in the 6 different environments.}
\end{figure*}

\section{Grain Motions arising from hydro-drag}

In hydrodynamic turbulence, the grain motions are caused by the frictional
interaction with the gas. On large scales grains are coupled with the
ambient gas, and the fluctuating gas motions mostly cause an overall
advection of the grains with the gas (Draine 1985). At small scales
grains are decoupled. The largest velocity difference occurs on the
largest scale where grains are still decoupled. Thus the characteristic
velocity of a grain with respect to the gas corresponds to the velocity
dispersion of the turbulence on the time scale $t_d$. In the MHD
case, the charged grains are subject to electromagnetic forces.
If $\tauLar > t_d$,
the grain does not feel the
magnetic field. Otherwise, if $\tauLar< t_d$, grain perpendicular motions are constrained by the magnetic
field.

As Alfv\'{e}nic turbulence is anisotropic, it is convenient to consider
separately grain motions parallel and perpendicular to the magnetic
field. The motions perpendicular to the magnetic field are influenced
by Alfv\'{e}n modes, while those parallel to the magnetic field are
subjected to the magnetosonic modes. According to $\S2$, the perpendicular
velocity field scales as $v_{k}\approx V\left(\tau_{k}/\tau_{max}\right)^{1/2},$
where $\tau_{max}=L/V$ is the time-scale on the injection scale.

If the Larmor time $\tauLar < \tau_d$, grain perpendicular motions are constrained
by the magnetic field. In this case, grains have a velocity dispersion
determined by the turbulence eddy whose turnover period is $\sim\tauLar$,
while grains move with the magnetic field on longer time scales. Since
the turbulence velocity grows with the eddy size, the largest velocity
difference occurs on the largest scale where grains are still decoupled.
Thus, following the approach in Draine (1985), we can estimate the
characteristic grain velocity relative to the fluid as the velocity
of the eddy with a turnover time equal to $\tauLar$, 

\begin{eqnarray}
v_{\perp}(a) & = & {\frac{V^{3/2}}{L^{1/2}}}(\rho_{gr})^{1/2}\left({\frac{8\pi^{2}c}{3qB}}\right)^{1/2}a^{3/2}\nonumber \\
 & = & 0.9\times10^{5}\cm\s^{-1}(\Vfive a_{5})^{3/2}/(Z\Lten B_{\mu})^{1/2},\label{dvperp}\end{eqnarray}
 in which $\Vfive =V/10^{5}\cm\s^{-1}$,
$a_{5}=a/10^{-5}$cm, $Z=q/e$,
$\Lten =L/10$pc, $B_{\mu}=B/1\mu$G, and $\rho_{gr}$ is the mass density
of grain. We adopt $\rho_{gr}=3.8$g~cm$^{-3}$ for silicate grains
and $\rho_{gr}=2$gcm$^{-3}$ for carbonaceous grains. 

Grain motions parallel to the magnetic field are induced by the compressive component of the slow mode with $v_{\parallel}\approx V\times\tau_{k}/\tau_{max}$%
\footnote{We assume that turbulence is driven isotropically at the injection
scale $L$. %
}. For grain motions parallel to the magnetic field the Larmor precession
is unimportant and the gas-grain coupling takes place on the translational
drag time $t_d$. The drag time due to collisions with atoms
$t'_d=(a\rho_{gr}/n_{n})(\pi/8m_{n}k_{B}T)^{1/2}$,
where $a$ is the grain size, $m_{n}$ is the mass of gas species,
$T$ is the temperature,
is essentially the time for collision with the mass of gas equal to
the mass of grain.
The ion-grain cross-section due to long-range
Coulomb force is larger than the atom-grain cross-section . For subsonic motions, the effective drag time $t_d^0=t'_d/\alpha$, where (Draine \& Salpeter 1979)

\begin{eqnarray}
\alpha&=&\left[1+\frac{n_{\rm H}}{2n_n}\sum_{i}x_i\left(\frac{m_{i}}{m_n}\right)^{1/2}\sum_{Z}f_Z\left(\frac{Ze^{2}}{ak_BT}\right)^{2}\right. \nonumber \\
& &\left. \ln\left[\frac{3(k_B
T)^{3/2}}{2e^{3}|Z|(\pi xn_{\rm H})^{1/2}}\right]\right]^{-1},
\end{eqnarray}
 where $x_{i}$ is the abundance, relative to hydrogen, of ion $i$ with mass $m_{i}$, $x=\sum_i x_i$, and $f_{Z}$ is the probability of the grain being in charge state $Z$.

The characteristic velocity of grain motions along the magnetic field
is approximately equal to the parallel turbulent velocity of eddies
with turnover time equal to $t_d$
\begin{eqnarray}
\label{eq:v||}
v_{\parallel}(a) & = & \alpha^{-1}{\frac{V^{2}}{L}}\left(\frac{\rho_{gr}}{4n_{n}}\right)({\frac{2\pi}{m_{n}k_{B}T}})^{1/2}a\nonumber \\
 & = & (3.8\times10^{5}\cm/s)\alpha^{-1} \Vfive^{2}a_{5}/(n\Lten T_{100}^{1/2}),\end{eqnarray}
where $T_{100}=T/100$K. Eq.\ (\ref{eq:v||}) is valid for subsonic motion, 
$v < (k_B T/m_n)^{1/2}$.

When the grain velocity 
$v$ relative to the gas becomes supersonic (Purcell 1969), the
gas drag time $t_d \approx t_d^0/(1+(9\pi/128)v^2/C_s^2)^{1/2}$.

When $\tauLar>t_d$, grains are no longer tied to the magnetic
field. Since at a given scale, the largest velocity dispersion is
perpendicular to the magnetic field direction, the velocity gradient
over the grain mean free path is maximal in the direction perpendicular
to the magnetic field direction. The corresponding scaling is analogous
to the hydrodynamic case, which was discussed in Draine (1985): 

\begin{eqnarray}
v(a) & = & {\frac{V^{3/2}}{L^{1/2}}}t_d^{1/2}\nonumber \\
 & = & \alpha^{-1/2}{\frac{V^{3/2}}{L^{1/2}}}\left(\frac{\rho_{gr}}{4n_{n}}\right)^{1/2}\left({\frac{2\pi}{m_{n}k_{B}T}}\right)^{1/4}a^{1/2}.\label{HD}\end{eqnarray}

It is easy to see that the grain motions get modified when the damping time scale of the turbulence $\tau_{c}$ is longer than either $t_d$
or $\tauLar$. In this case, a grain samples only
a part of the eddy before gaining the velocity of the ambient gas.
In a turbulent medium, the shear rate $dv/dl$ increases with the decrease
of eddy size. Thus for $\tau_{c}>\max\{ t_d,\tauLar\}$, these
smallest available eddies are the most important for grain acceleration.
Consider first the perpendicular motions. If $v_{c}$ is the velocity
of the critically damped eddy, the distance traveled by the grain
is $\bigtriangleup l\sim v_{c}\times \min\lbrace t_d,\tauLar\rbrace$.
The shear rate $dv/dl$ perpendicular to the magnetic field is $\tau_{k}^{-1}$.
Thus the grain experiences the velocity difference in the direction
perpendicular to the magnetic field

\begin{equation}
v_{\perp}\sim\bigtriangleup l\times\frac{dv}{dl}\sim\frac{v_{c}}{\tau_{c}}\times\min\lbrace t_d,\tauLar\rbrace.\label{vperp'}\end{equation}
 
For the parallel motions, $\bigtriangleup l\sim v_{c}\times t_d$.
From the critical balance in the GS95 model $k_{\parallel}V_{A}\sim k_{\perp}v_{\perp}=\tau_{k}^{-1}$,
the largest shear rate along the magnetic field should be $dv/dl=v_{c}k_{\parallel}\sim v_{c}/(V_{A}\tau_{c})$.
Therefore, in the parallel direction, the grain experiences a velocity
difference $V_{A}/v_{c}$ times smaller, i.e.,

\[
v_{\parallel}\sim\frac{v_{c}^{2}}{V_{A}}\times\frac{t_d}{\tau_{c}}.\]

The velocity dispersion induced by the compressional motion associated
with the fast modes also causes motion relative 
to the ambient gas. The velocity fluctuation for fast modes scales
as $v_{k}\propto k^{-1/4}\propto\omega^{-1/4}$, where $\omega$ is
the frequency of fast modes. From similar considerations,
we know that grains get velocity dispersions during 
$\min\lbrace\tauLar,t_d\rbrace$,i.e. , 
$v\simeq V(\min\lbrace\tauLar,t_d\rbrace/\tau_{max})^{1/4}$.
Grains with $\min\lbrace\tauLar,t_d\rbrace<\tau_{c}$, have reduced velocities
$v\sim v_{c}\times\tauLar/\tau_{c}\sim V(\tauLar/\tau_{max})^{1/4}(\tauLar/\tau_{c})^{3/4}$,
where $v_{c}$ is the velocity of turbulence at the damping scale.
From the scaling, we see that the decoupling from fast modes always
brings larger velocity dispersions to grains than Alfv\'{e}n modes
($v_{k}\propto k^{-1/3}$) except for the situation when Alfv\'{e}n modes
dominate MHD turbulence. The velocity fluctuations associated with
fast modes are always in the direction perpendicular to $\mathbf{B}$
in low $\beta$ media (see Appendix D). Thus the grain velocities
are also perpendicular to $\mathbf{B}$. In high $\beta$ media,
grains can have velocity dispersion in any direction as the velocity
dispersions of fast modes are longitudinal, i.e., along $\mathbf{k}$.

\section{Acceleration of Grains by Gyroresonance}

Gyroresonance acceleration of charged grains by a spectrum
of MHD waves decomposed into incompressive Alfv\'{e}nic,
and compressive fast and slow modes (see CL02) was first described
in YL02. The resonance happens when $\omega-k_{\parallel}v\mu=n\Omega$,
($n=0,\pm1,\pm2...$), where $\omega$ is the wave frequency, $k_{\parallel}$
is the parallel component of wave vector ${\textbf{k}}$ along the
magnetic field, $v$ is the grain velocity, $\mu$ is the cosine of
the grain pitch angle relative to the magnetic field, and $\Omega$ is
the Larmor frequency of the grain. There are
two main types of resonant interactions: gyroresonance acceleration
and transit acceleration. Transit acceleration ($n=0$) requires longitudinal
motions and only operates with compressive modes. It happens when
$k_{\parallel}v\mu=\omega$, which requires particle speed to be super-Alfv\'enic $v >V_f \ge V_A$. Although this condition is partially relieved owing to resonance broadening (see Yan \& Lazarian 2004), transit acceleration of low speed grains is marginal
because sub-Alfv\'enic particles can hardly catch up with the moving magnetic mirror.

How can we understand grain gyroresonance? Gyroresonance occurs when
the Doppler shifted frequency of the wave in the grain's guiding center
rest frame $\omega_{gc}=\omega-k_{\parallel}v\mu$ is a multiple of
the grain gyrofrequency.
For low speed grains, we only need to consider the resonance at $n=1.$ The gyroresonance
changes both the direction and absolute value of the grain's momentum
(i.e., scatters and accelerates the grain). The efficiency of the 
two processes for charged grains can be described
by the Fokker-Planck coefficients $D_{\mu\mu}$ and $D_{pp}/p^{2}$,
where $p$ is the grain momentum. The ratio of the two rates depends
on the ratio of the grain velocity and the Alfv\'{e}n speed and pitch
angle, $p^{2}D_{\mu\mu}/D_{pp}=[(v\zeta/V_{A})+\mu]^{2}$, where $\zeta=1$
for Alfv\'{e}n waves and $\zeta=k_{\parallel}/k$ for fast modes
(see Appendix B). We see that the scattering is less efficient for
sub-Alfv\'{e}nic grains unless most grains move parallel to the magnetic
field. We shall show later that as the result of acceleration, $\mu$
will tend to 0. Therefore in the zeroth order approximation, we ignore
the effect of scattering and assume that the pitch angle cosine $\mu$
does not change while a grain is accelerated. In this case, the Fokker-Planck
equation, which describes the diffusion of grains in momentum
space, can be simplified (see Pryadko \& Petrosian 1997):

\begin{equation}
\frac{\partial f^{\mu}}{\partial t}+v\mu\frac{\partial f^{\mu}}{\partial z}=\frac{1}{p^{2}}\frac{\partial}{\partial p}p^{2}D_{pp}(\mu)\frac{\partial f^{\mu}}{\partial p},\label{eq:Fokker}\end{equation}
where $f$ is the distribution function. Apart from acceleration,
a grain is subjected to gaseous friction. Thus we describe the stochastic
acceleration by the Brownian motion equation:

\begin{equation}
m\frac{dv}{dt}=-\frac{v}{S}+Y,\label{eq:moment}\end{equation}
 where $m$ is the grain mass, $Y$ is the stochastic acceleration
force, $S=t_d/m$ is the mobility coefficient. 

If we multiply equation(\ref{eq:moment}) by $v$, and take the ensemble average, we obtain

\begin{equation}
m\frac{d\langle v^{2}\rangle}{dt}=-\frac{\langle v^{2}\rangle}{S}+\langle\dot{\epsilon}^{\mu}\rangle.
\label{diff}
\end{equation}
The steady solution is achieved when the derivative on
the left-hand is zero. Following an approach similar to that in Melrose (1980), we can get from Eq.(\ref{eq:Fokker}) the energy gain rate for the grain
with pitch angle $\mu$

\begin{equation}
\langle
\dot{\epsilon}^{\mu}\rangle
=\frac{1}{p^{2}}\frac{\partial}{\partial p}(vp^{2}D_{pp}(\mu)).\label{epsilon}\end{equation}
 The Fokker-Planck coefficient $D_{pp}(\mu)$ is calculated below.

YL03 employed quasi-linear theory (QLT) to obtain $D_{pp}(\mu)$
(see also YL02 and Appendix B)
\footnote{Usually the real part is taken of the integral. However, we show in Appendix B 
that the integrand is real.}:

\begin{eqnarray}
D_{pp}(\mu) & = & {\frac{\pi\Omega^{2}(1-\mu^{2})p^{2}V_{A}^{2}}{2v^{2}}}
\Big\{
\int dk^{3}\frac{\tau_{k}^{-1}}{\tau_{k}^{-2}+(\omega-k_{\parallel}v\mu-\Omega)^{2}}\nonumber \\
 &  & 
\Big[
(J_{2}^{2}({\frac{k_{\perp}v_{\perp}}{\Omega}})M_{\mathcal{RR}}({\mathbf{k}}))+J_{0}^{2}({\frac{k_{\perp}v_{\perp}}{\Omega}})M_{\mathcal{LL}}({\mathbf{k}}))\nonumber \\
 & - & J_{2}({\frac{k_{\perp}v_{\perp}}{\Omega}})J_{0}({\frac{k_{\perp}v_{\perp}}{\Omega}})\nonumber \\
 &  & (e^{i2\phi}M_{\mathcal{RL}}({\mathbf{k}})+e^{-i2\phi}M_{\mathcal{LR}}({\mathbf{k}}))
\Big]\Big\}.
\label{spep}
\end{eqnarray}
However, we should not integrate
over all $k$ because the contribution from large scales is spurious (see discussion in YL02). This contribution stems from the fact that in QLT,
an unperturbed grain orbit is assumed, which results in non-conservation
of the adiabatic invariant $\xi=mv_{\perp}^{2}/2B_{0}$. Noticing
that the adiabatic invariant is conserved when the electromagnetic
field varies on a time scale longer than $\Omega^{-1}$, we truncate
our integral range, namely, integrate from $k_{res}$ instead of the
injection scale $L^{-1}$. For Alfv\'{e}nic turbulence $\omega=|k_{\parallel}|V_{A}$,
the resonant scale corresponds to $|k_{\parallel,res}|=\Omega/|V_{A}-v\mu|.$
For fast modes, the resonant scale is $k_{res}=\Omega/|V_{f}-v\mu\cos\theta|$,
where $\cos\theta=k_{\parallel}/k$. The upper limit of the integral
$k_{c}$ is set by the dissipation of the MHD turbulence, which varies
with the medium. 

Integrating from $k_{res}$ to $k_{c}$, we obtain from Eq.(\ref{spep}) and (\ref{epsilon}) the energy gain rate $\dot{\epsilon}$ as a function
of $v$ and $\mu$. Then with $\dot{\epsilon}$
known, we estimate the grain acceleration. Solving Eq.(\ref{diff})
iteratively, we obtain the grain velocity as a function of time.
We check that the grain velocities converge to a constant value after the drag time. As $\dot{\epsilon}$ increases with pitch
angle, grains gain the maximum velocities perpendicular to the magnetic
field and therefore the averaged $\mu$ decreases. This is understandable
since the electric field, which accelerates the grain, is in the direction
perpendicular to the magnetic field.

\section{Grain motions in the ISM }

Here we apply our results to various idealized phases of the interstellar medium.

\begin{table*}
\begin{tabular}{|c|c|c|c|c|c|c|}
\hline 
 ISM&
 CNM&
 WNM&
 WIM&
 MC&
 DC1&
 DC2\tabularnewline
\hline
\hline
T(K)&
 100&
 6000&
 8000&
 25&
 \multicolumn{2}{|c|}{10}\tabularnewline
\hline
$\nH$(cm$^{-3}$)&
 30&
 0.3&
 0.1&
 300&
 \multicolumn{2}{|c|}{$10^4$}\tabularnewline
\hline
$n_e$(cm$^{-3}$)&
0.03&
0.03&
0.0991&
0.03&
0.01&
0.001\tabularnewline
\hline
$G_{UV}$&
1&
1&
1&
0.1&
0.01&
0.001\tabularnewline
\hline
B($\mu$G)&
6&
5.8&
3.35&
11&
\multicolumn{2}{|c|}{80}\tabularnewline
\hline
L(pc)&
0.64*&
100&
100&
1&
\multicolumn{2}{|c|}{1}\tabularnewline
\hline
$V=V_A$(km/s)&
2*&
20&
20&
1.2&
\multicolumn{2}{|c|}{1.5}\tabularnewline
\hline
damping&
  neutral-ion&
 neutral-ion&
  collisional&
neutral-ion&
\multicolumn{2}{|c|}{ion-neutral decoupling}\tabularnewline
\hline
$k_c(\cm^{-1})$&
$7\times 10^{-15}$&
$4\times 10^{-17}$&
$NA$&
$4.5\times 10^{-14}$&
$5.3\times 10^{-15}$&
$5.3\times 10^{-17}$\tabularnewline
\hline
\end{tabular}\centering

\caption{The parameters of idealized ISM phases and relevant damping. 
Among them, $\nH$ is the number density of H, $n_e$ is the number density of 
electron, $G_{UV}$ is the UV intensity scale factor, 
L is the injection scale of fast modes, V is the injection velocity. The
dominant damping mechanisms for fast modes are given with the corresponding damping timescale $\tau_c$. CNM=cold neutral medium, WNM=warm neutral
medium, WIM=warm ionized medium, MC=molecular cloud, DC=dark cloud. * See text for the explanation of smaller L and V for CNM.}

\end{table*}

First consider a typical cold neutral medium (CNM), $T=100$K, $\nH=30$cm$^{-3},$
$n_{e}=0.045$cm$^{-3}$, $B=6\mu$G, with corresponding $v_A=2$~km/s and $\beta\sim 0.4$.

Our treatment of
MHD turbulence requires
that fluid velocities are smaller than
the Alfv\'{e}n speed%
\footnote{Otherwise magnetic field is not dynamically important. Turbulence is essentially hydrodynamic. See the following discussions.}. Therefore
we assume that the injection of energy happens at an effective injection
scale $L$ where equipartition between magnetic and kinetic energies, i.e., $V=V_{A}$, occurs.
This effective injection scale can be different from the actual scale at which energy is injected. For instance, if we assume that the velocity dispersion at the scale
$l=10$pc is 5km/s, this means that turbulence in the
CNM is super-Alfv\'{e}nic
at this scale. The turbulence then follows a
hydrodynamic cascade down to a scale $L\approx0.64$pc 
where the turbulent velocity becomes equal to 
$V_A=2$km/s; we identify this as the effective
``injection scale'' $L$\footnote{
This picture is not self-consistent as we expect to have turbulent
generation of magnetic field which will bring kinetic and magnetic
energy to equipartition at the injection scale (see arguments in Cho,
Lazarian \& Vishniac 2002a), but the idea of super-Alfv\'{e}nic turbulence percolates
in the literature (Boldyrev, Nordlund \& Padoan 2002).%
}with injection velocity $V=V_{A}=2$km/s. Alternatively, it is possible that the turbulence
at large scales proceeds in tenuous warm media with Alfv\'{e}n speed
larger or equal to 5km/s. Nevertheless, the
statistics of fast modes will not be changed in 
the CNM.

In partially ionized media, the damping is dominated by the viscosity
arising from neutrals. The turbulence is assumed damped%
\footnote{Thus we ignore the effect of slowly evolving magnetic structures associated with a recent reported new regime of turbulence below the viscous
damping cutoff (Cho, Lazarian, \& Vishniac 2002c, Lazarian, Vishniac, \& Cho 2004).%
} when its cascading time scale $\tau_{k}=t_{damp}$; this defines
the cutoff scale $k_{\parallel,c}=4\times10^{-16}$cm$^{-1}$ for
Alfv\'{e}n modes and $k_{c}=7\times10^{-15}$cm$^{-1}$ for fast
modes (see Appendix A). Assuming that the grain velocities are smaller
than the phase speed of fast modes, we find that the prerequisite
for gyroresonance $k_{c}>k_{res}$ is the same as $\tauLar>\tau_{c},$
the condition for effective hydro drag (see Fig.1). For a silicate
grain, the critical grain size $a_c \approx 4\times10^{-6}$cm for fast modes
and $a_c\approx 10^{-5}$cm for Alfv\'{e}n modes.
Grains smaller than the critical size are not effectively
accelerated by the corresponding turbulent mode.

The CNM is a low $\beta$ medium, so the correlation tensors for the low
$\beta$ case are applied. As has been discussed in YL02, the interactions with Alfv\'{e}n modes are less efficient than those with fast modes because of the
anisotropy of Alfv\'{e}n modes. Thus we shall consider
only fast modes for later calculations. The gyroresonance with fast
modes in the CNM can accelerate grains to supersonic
velocities. $\dot{\epsilon}^{\mu}$ increases with pitch angle. If we average $\dot{\epsilon}$ over $\mu$, we will get mean velocities which are smaller than the maximum values by less than $20\%$. 
In Fig.1 we plot the velocity of grains with pitch
angle equal to $90^{o}$ as a function of grain size since all the
mechanisms preferentially accelerate grains in this direction.

\begin{figure}
\includegraphics[width=0.49\columnwidth]{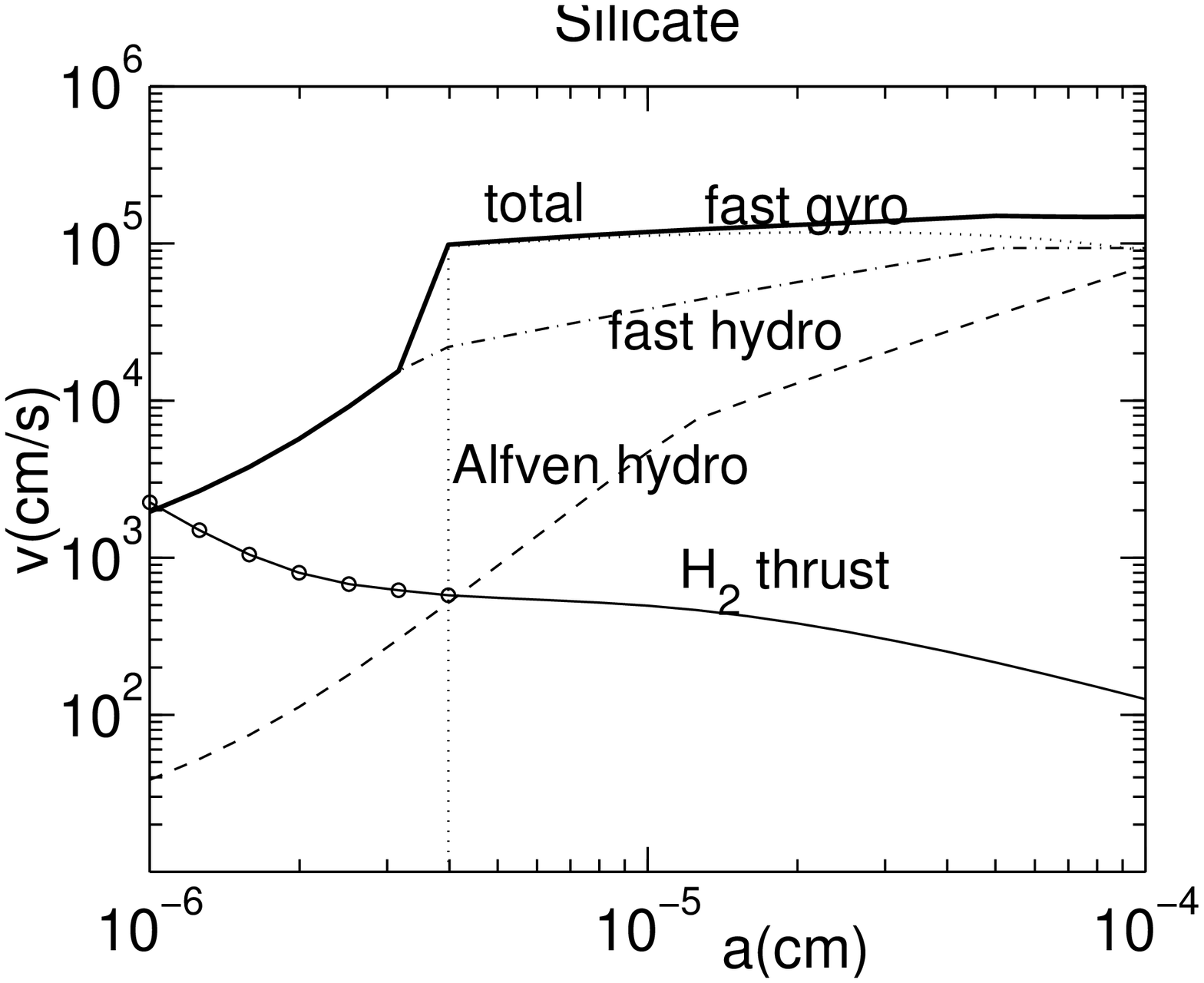} \hfil
\includegraphics[width=0.49\columnwidth]{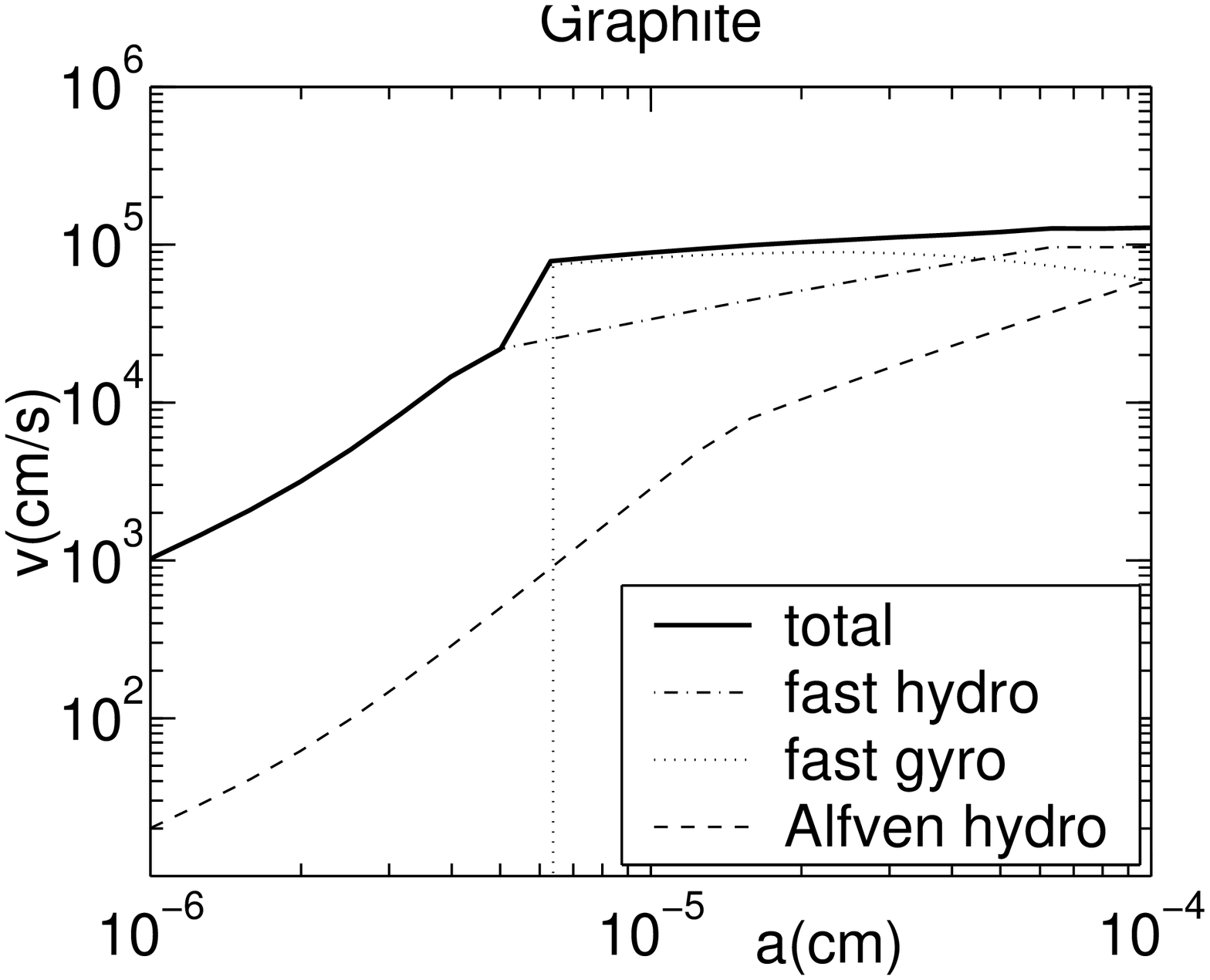} 

\caption{Relative velocities as a function of radii (solid line) in CNM, \emph{left}:
for silicate grains, \emph{right}: for graphite grains. The dotted
lines represent the gyroresonance with fast modes. Gyroresonance
works only for large grains owing to the cutoff by
viscous damping. The cutoff scales for fast and Alfv\'{e}n modes
are different due to their different scalings and the anisotropy of
Alfv\'{e}n modes. The dashed lines are the result from hydro drag by
Alfv\'{e}n modes (see LY02), the dashdot lines represent the hydro
drag by fast modes. Contributions from different mechanisms are
approximately additive in squares, i.e., $v_{tot}^{2}=\Sigma_{i}v_{i}^{2}$ (heavy solid lines).
The grain velocity driven by H$_{2}$ formation (dashdot line) is plotted
to illustrate the issue of grain segregation in CNM (see text). The
part marked by open circles is nonphysical because thermal flipping
is not taken into account.}

\end{figure}

For the warm neutral medium (WNM), $T=6000$K, $n_{H}=0.3$cm$^{-3}$,
$n_{e}=0.03$cm$^{-3}$, $B=5.8\mu$G. Assume the velocity dispersion
$V=V_{A}=20$km/s at the injection scale $L=100$pc. Turbulence is mainly subjected to neutral-ion
damping. The fast modes are cut off at 
$k_{c}=4\times10^{-17}\cm^{-1}$
(see Appendix A). Comparing $k_{c}$ with $k_{res}$, we find
$a_c\approx2\times10^{-5}$cm for silicate grains.
The WNM has $\beta\approx0.25$, so we use the tensor given in Eq.(\ref{fast})
for fast modes. Integrating from $k_{res}$ to $k_{c}$ and solving
Eq.(\ref{diff}), we obtain grain velocities. The maximum values
are shown in Fig.2. We see large grains
can be accelerated to supersonic speeds%
\footnote{Unlike hydro-drag the gyroresonance can potentially accelerate grains
to velocities much higher than the velocity of turbulent motions.
For typical ISM conditions this, however, does not happen. %
}. The fact that these grains approach the Alfv\'{e}n speed makes our approximation
that acceleration dominates scattering less accurate, but the result
is correct within a factor of unity. Smaller grains are accelerated only by the hydro drag, which is
far less effective. 

\begin{figure}
\centering \leavevmode
\includegraphics[%
  width=0.49\columnwidth]{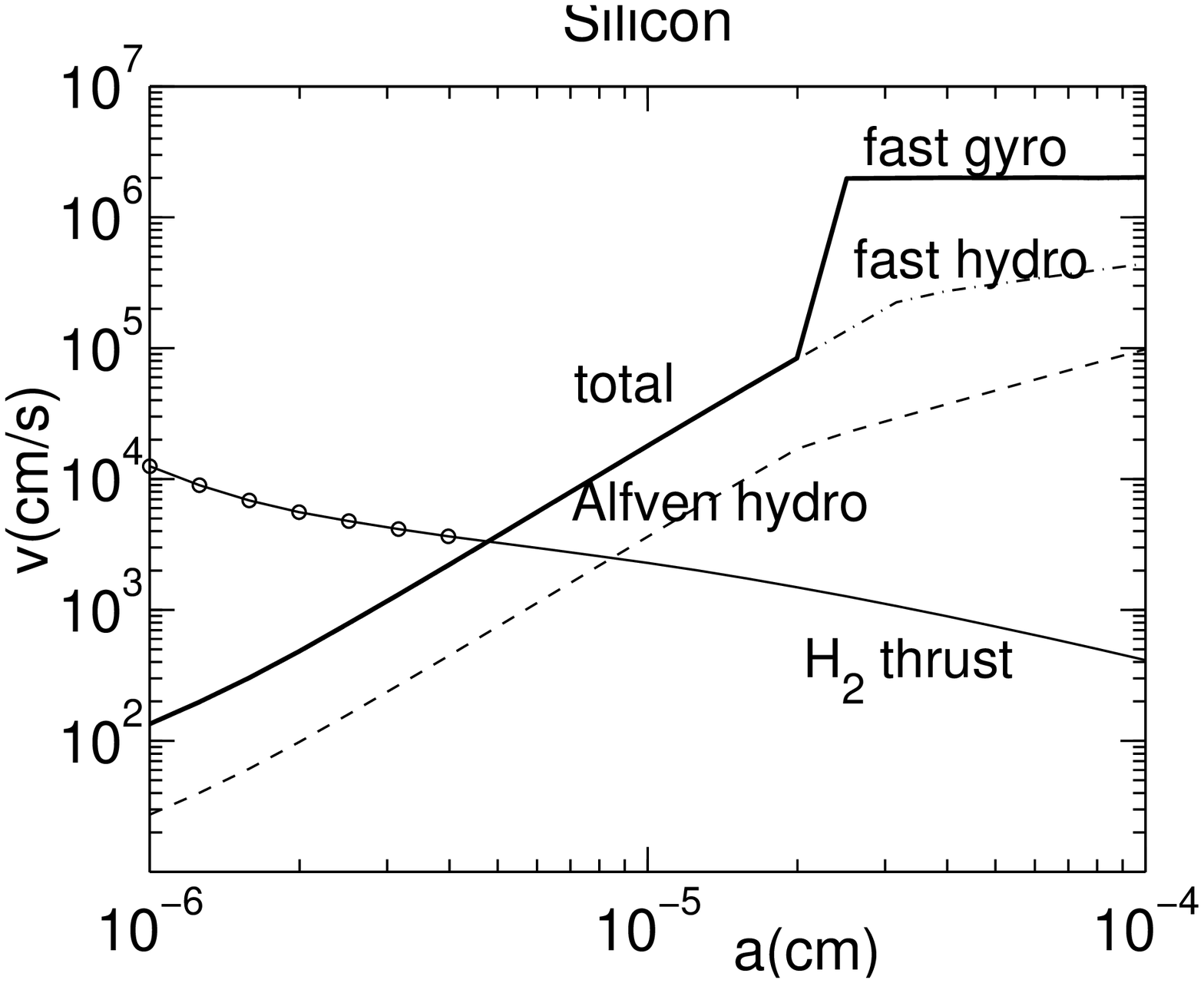} \hfil
 \includegraphics[%
  width=0.49\columnwidth]{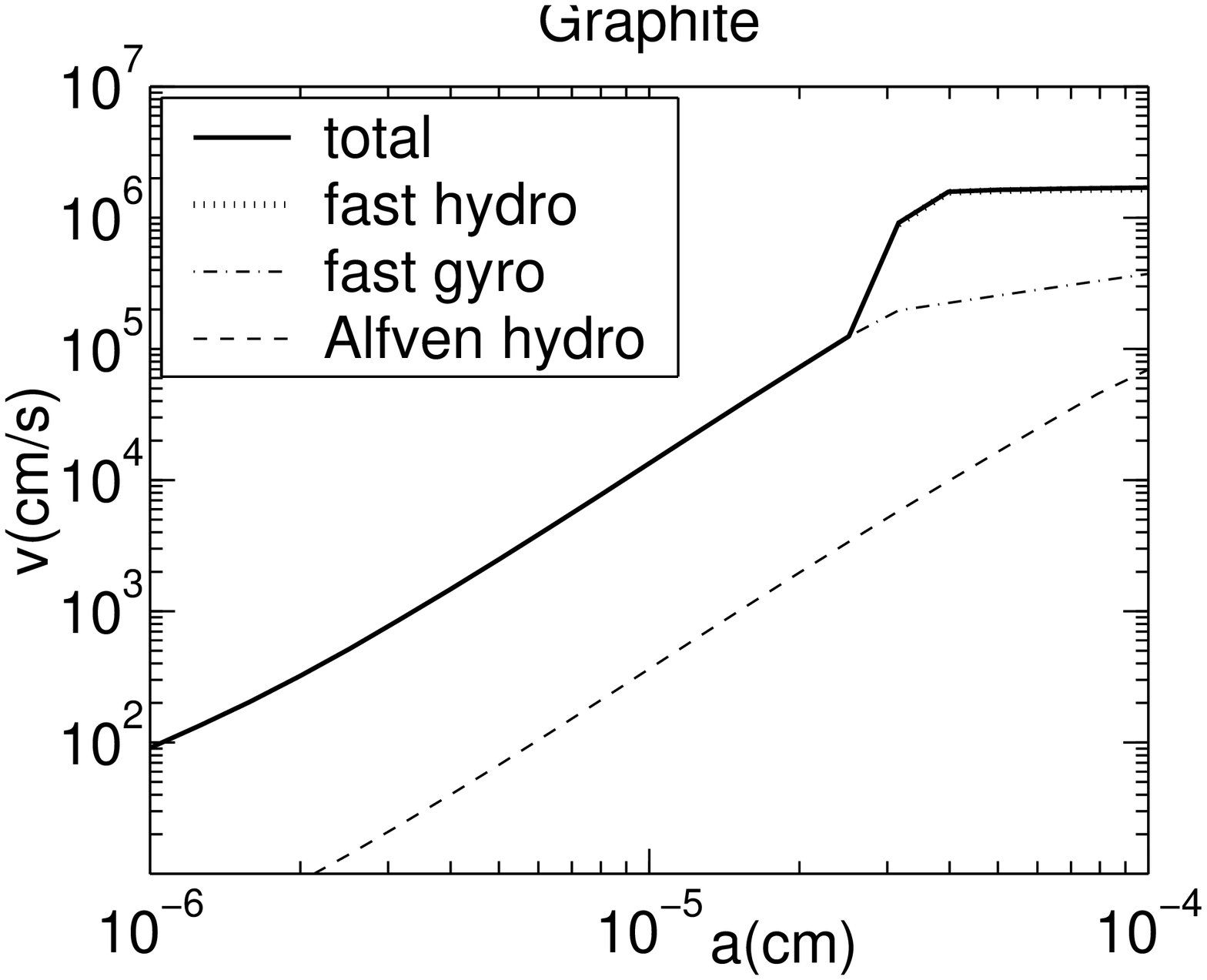}

\caption{Same as Fig2, but in the WNM. }
\end{figure}

The warm ionized medium (WIM) has $T=8000$K, 
$n_{e}=0.1\cm^{-3}$, $B=3.35\mu$G, with corresponding
$\beta\sim0.33$. The injection scale and speed are the same as in
the WNM. The WIM is fully ionized and in low $\beta$ regime. Fast
modes, in this case, are mainly affected by collisional damping. This damping increases with $\theta$,\footnote%
{This $\theta$ dependence makes the treatment of damping more complicated if taking into account field line wandering (see Yan \& Lazarian 2004). } and doesn't exist for parallel modes (see Appendix A). Thus there are always modes interacting with grains though the energy available is less at smaller scales. Following the same routine as above,
we get the grain velocities.  We see from Fig.4 the nonmonotonic variation of grain
velocity with the size. This arises from the fact that the charging
for grains in the WIM has a complex dependence on grain size (see also Fig.1a). 

\begin{figure}
\centering \leavevmode
\includegraphics[%
  width=0.49\columnwidth]{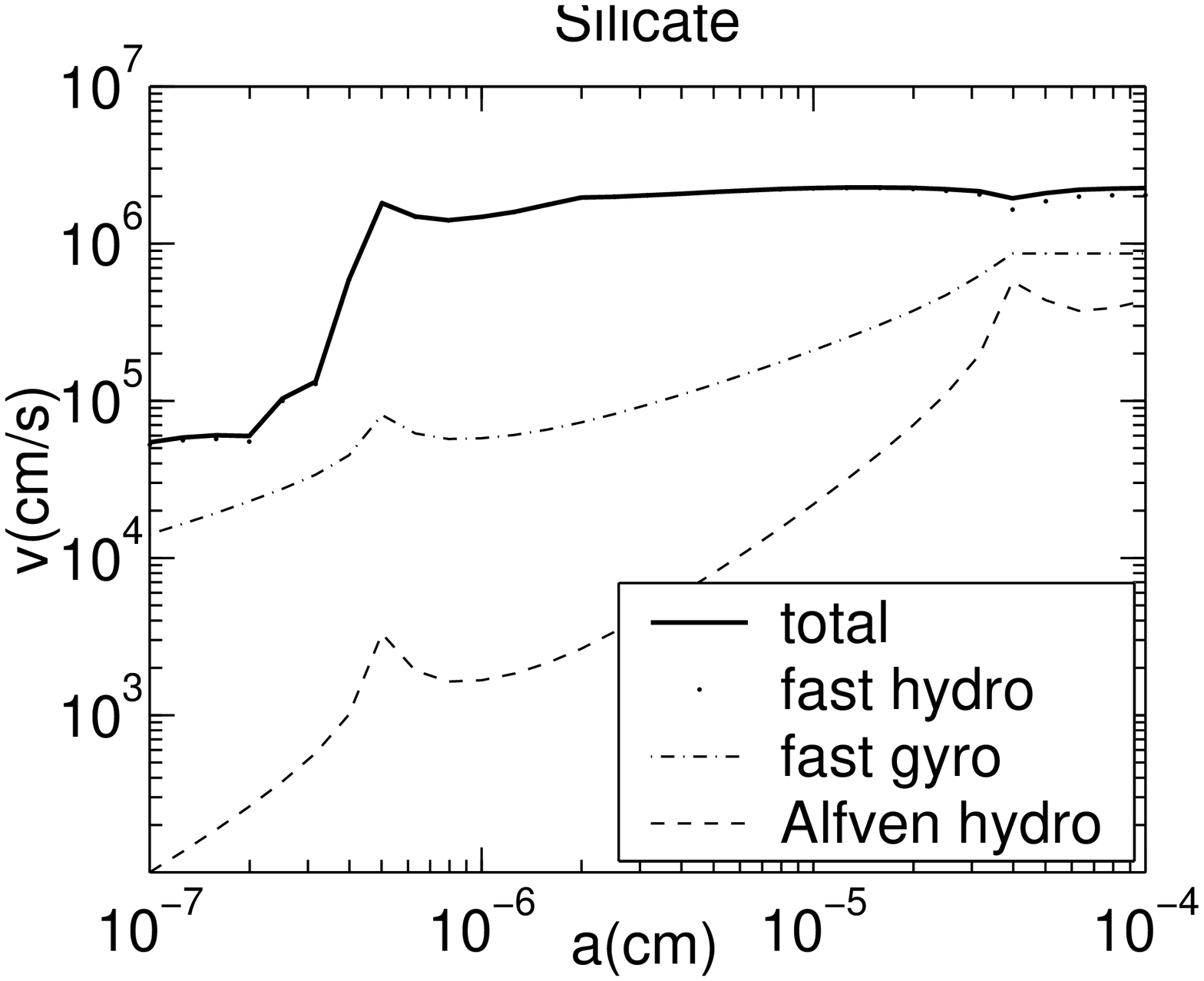} \hfil
 \includegraphics[%
  width=0.49\columnwidth]{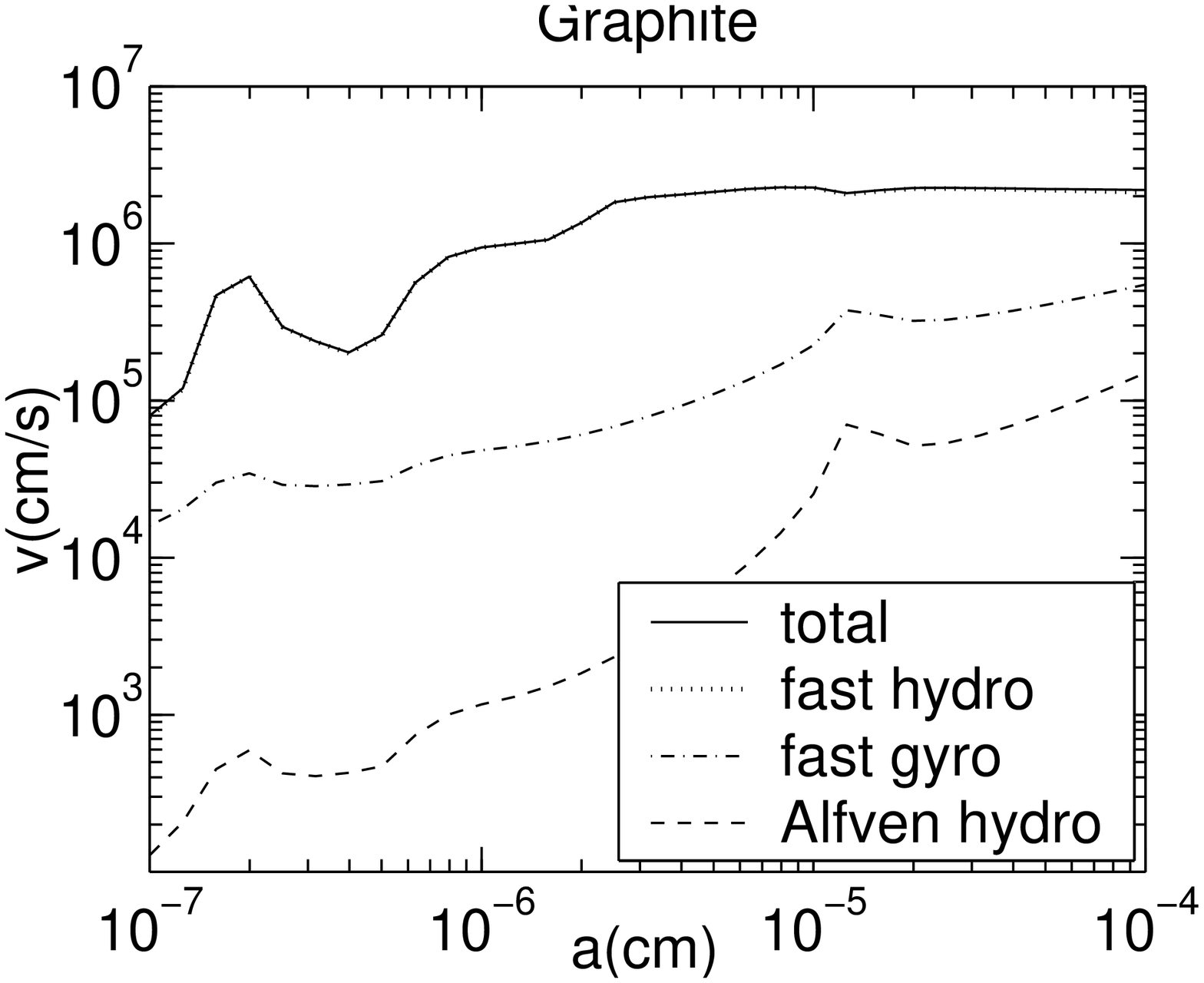}

\caption{Same as Fig.2, but in the WIM. The oscillations in these curves are due to the variation in charging
of grains.}
\end{figure}

Molecular cloud (MC) gas has $T=25$K,
$n_{\rm H}=300$cm$^{-3}$, and we adopt a
magnetic field strength $B=11\mu$G as suggested by 
observations (Crutcher 1999), corresponding to
Alfv\'en speed $V_A=1.2 {\rm \,km s\,}^{-1}$. The injection scale is
taken to be $L=1$pc and the injection velocity is $V=V_A$.
The damping scale (see Appendix A) of the turbulence is $k_{c}=4.5\times10^{-14}{\rm \,cm\,}^{-1}$, corresponding to resonant
scales of silicate grains with $a=8\times 10^{-7}$cm.
By following the same procedure, we obtain
the grain velocity distribution as shown in Fig.5.

We consider a typical dark cloud (DC) with $T=10$K,
$n_{\rm{H}}=10^{4}$cm$^{-3}$, and $B\sim 80\mu$G, 
corresponding to $V_A=1.5{\rm km\,s}^{-1}$.
The injection scale $L=1$pc 
and velocity $V=V_A$. The ionization in DC is so low that the fluid 
becomes decoupled in the middle of the cascade, where the ion-neutral 
collision rate $t_{ni}^{-1}$ per neutral is equal to the turbulence decay rate $\tau_k^{-1}$. Below this decoupling scale, neutrals will not follow ions and magnetic fields, and the
turbulence becomes hydrodynamic. In view of the uncertainty of the cosmic-ray ionization rate, we adopt two models DC1 and DC2 with electron densities $n_e=0.01\cm^{-3}$ and $n_e=0.001\cm^{-3}$. The grain charge distribution is the same for both DC1 and DC2, because we assume the same $G_{UV}/n_e$.
The main difference is the decoupling scale of the MHD cascade. Combining Eq.(\ref{cascade}) and Eq.(\ref{colli}), we can obtain the decoupling scales 
$k_c=5.3\times 10^{-15}\cm^{-1}$ for DC1 and $k_c=5.3\times 10^{-17}\cm^{-1}$ for DC2, corresponding to silicate grain size $a_c\approx3\times 10^{-6}$cm for DC1 and 
$a_c\approx2\times 10^{-5}\cm$ for DC2.
By following the same procedure, we obtain the grain velocity distribution as shown in Fig.~6 and Fig.~7.

\begin{figure}
\centering \leavevmode
\includegraphics[%
  width=0.49\columnwidth]{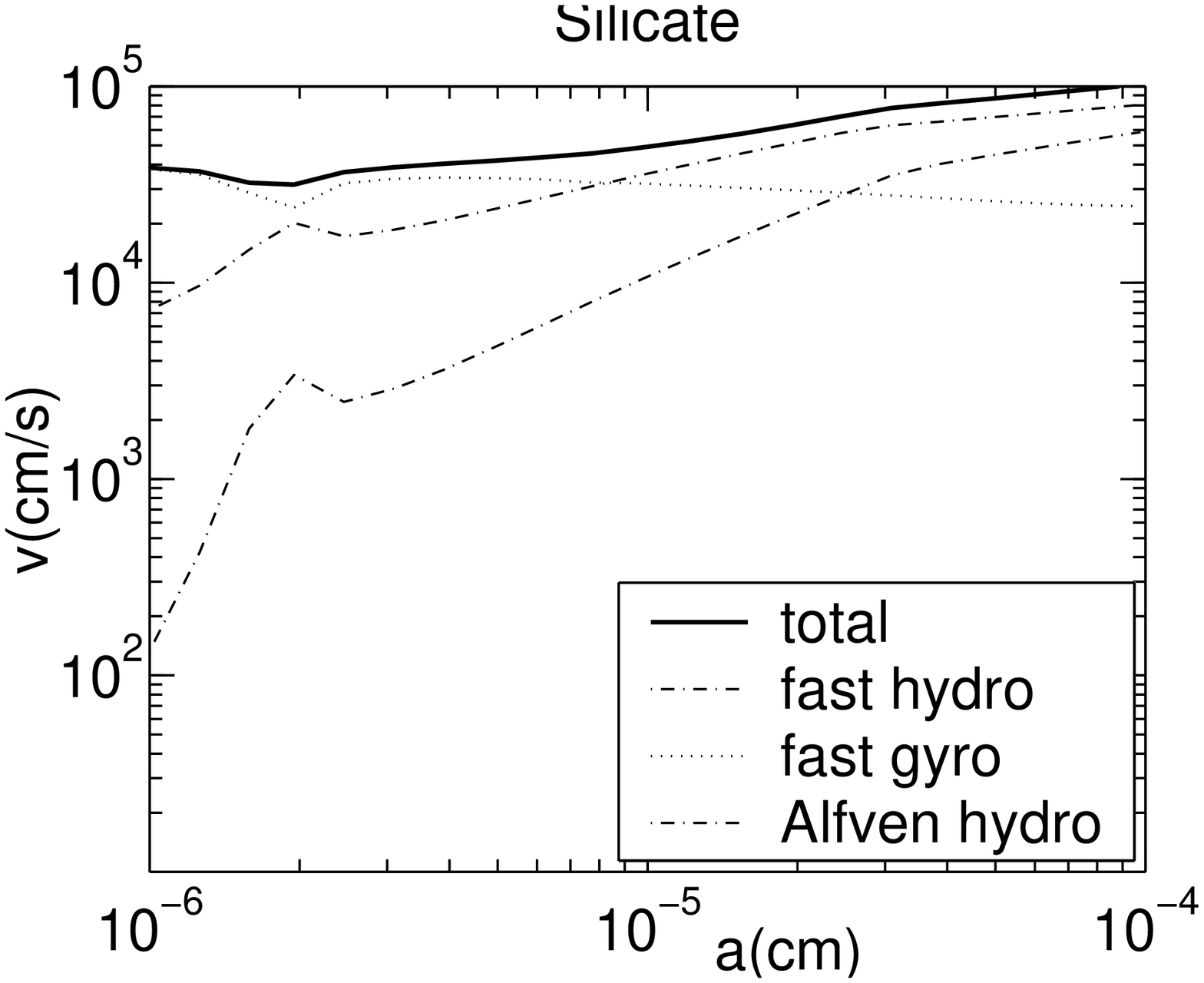} \hfil
 \includegraphics[%
  width=0.49\columnwidth]{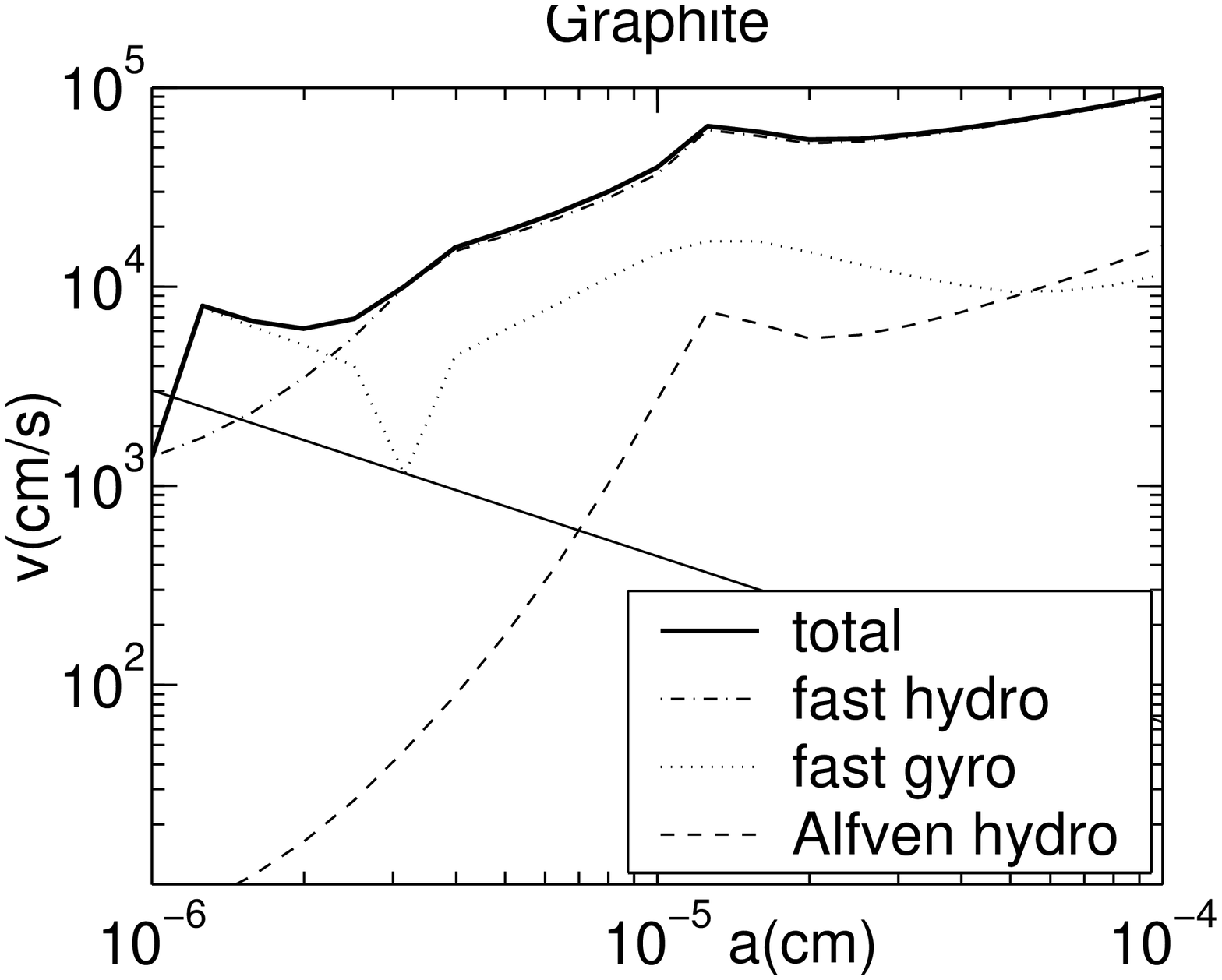} 
\caption{Same as Fig.4, but in the MC.}
\end{figure}

\begin{figure}
\centering \leavevmode
\includegraphics[%
  width=0.49\columnwidth]{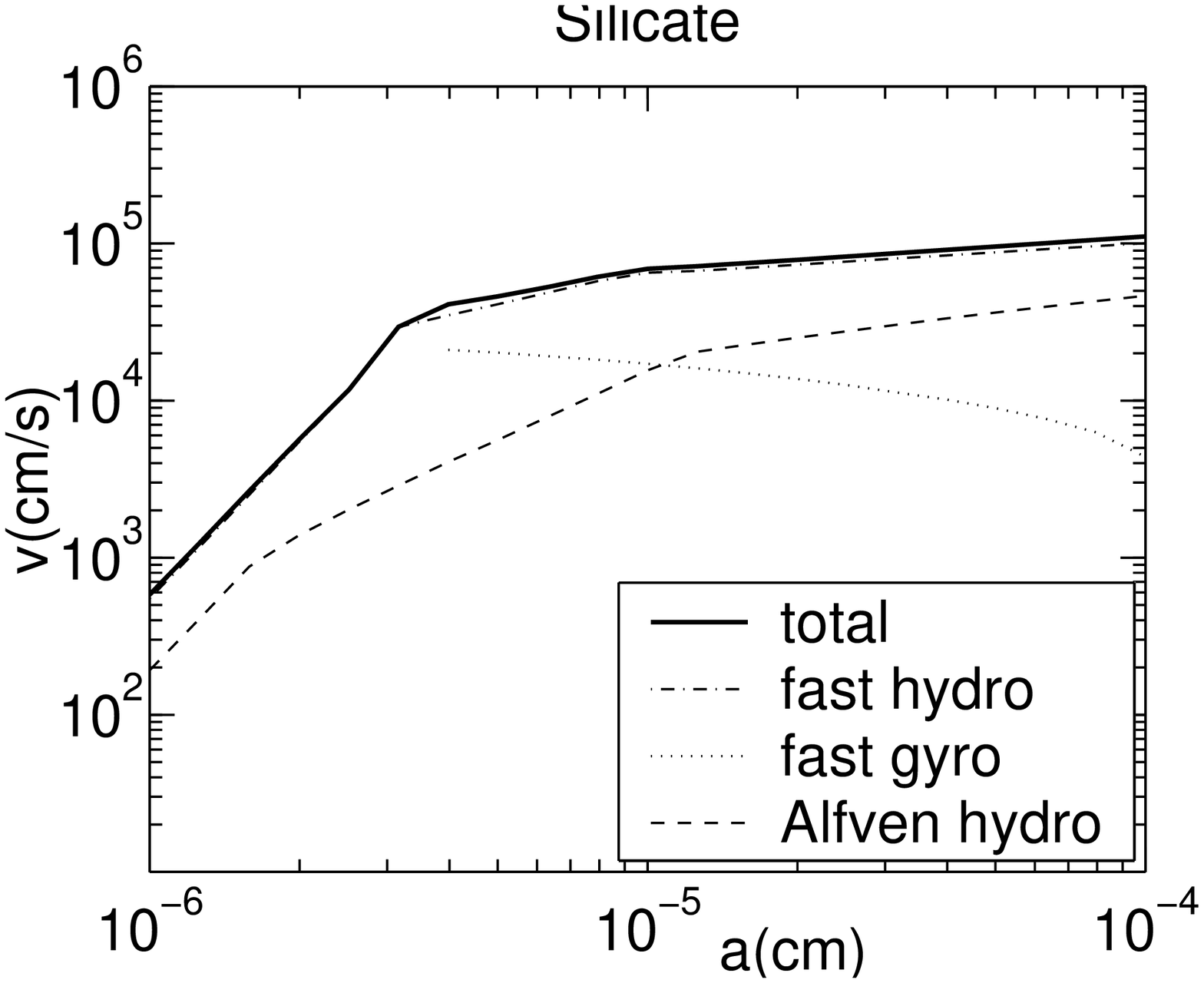} \hfil
 \includegraphics[%
  width=0.49\columnwidth]{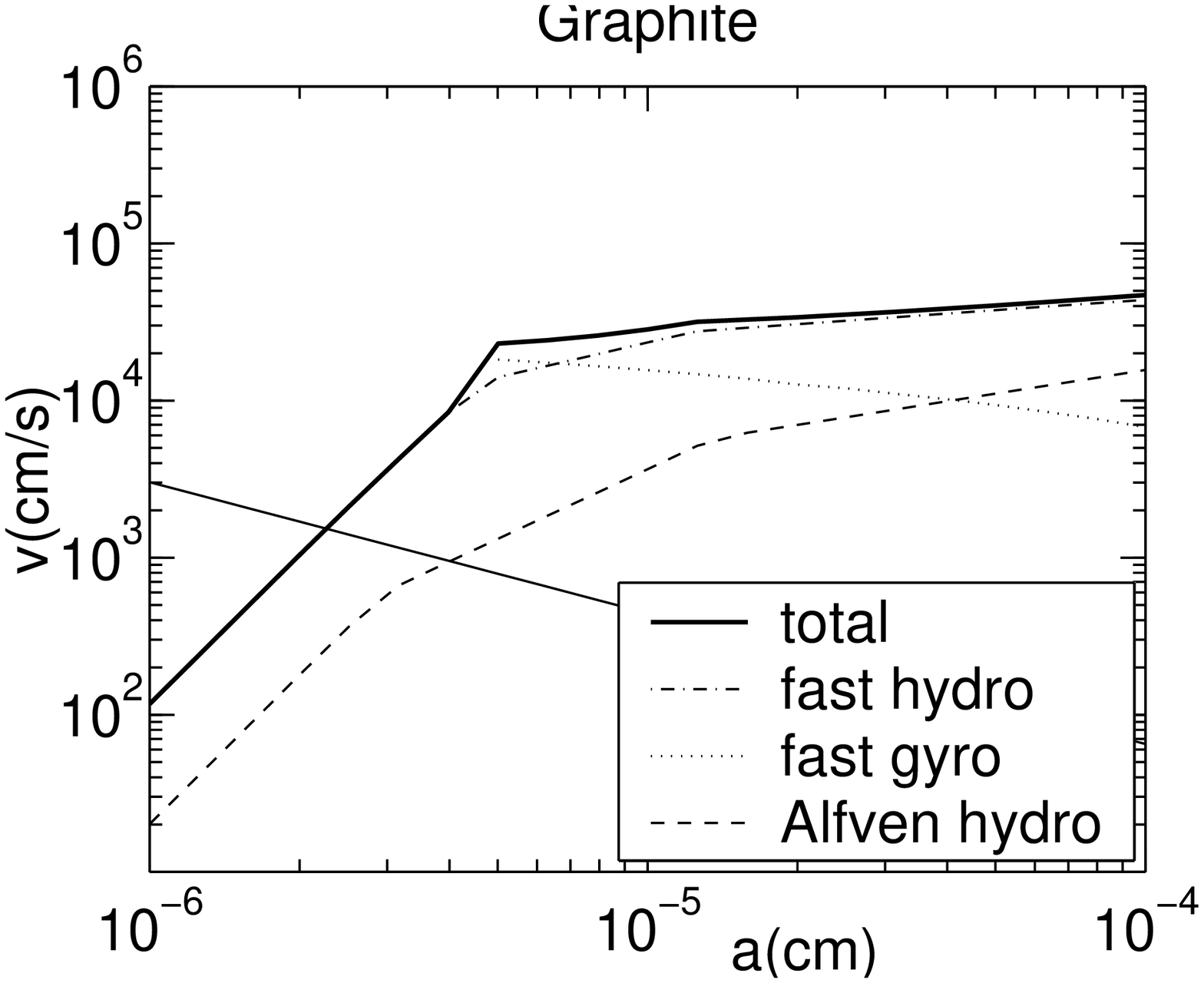}

\caption{Same as Fig.2, but in DC1.}
\end{figure}

\begin{figure}
\centering \leavevmode
\includegraphics[%
  width=0.49\columnwidth]{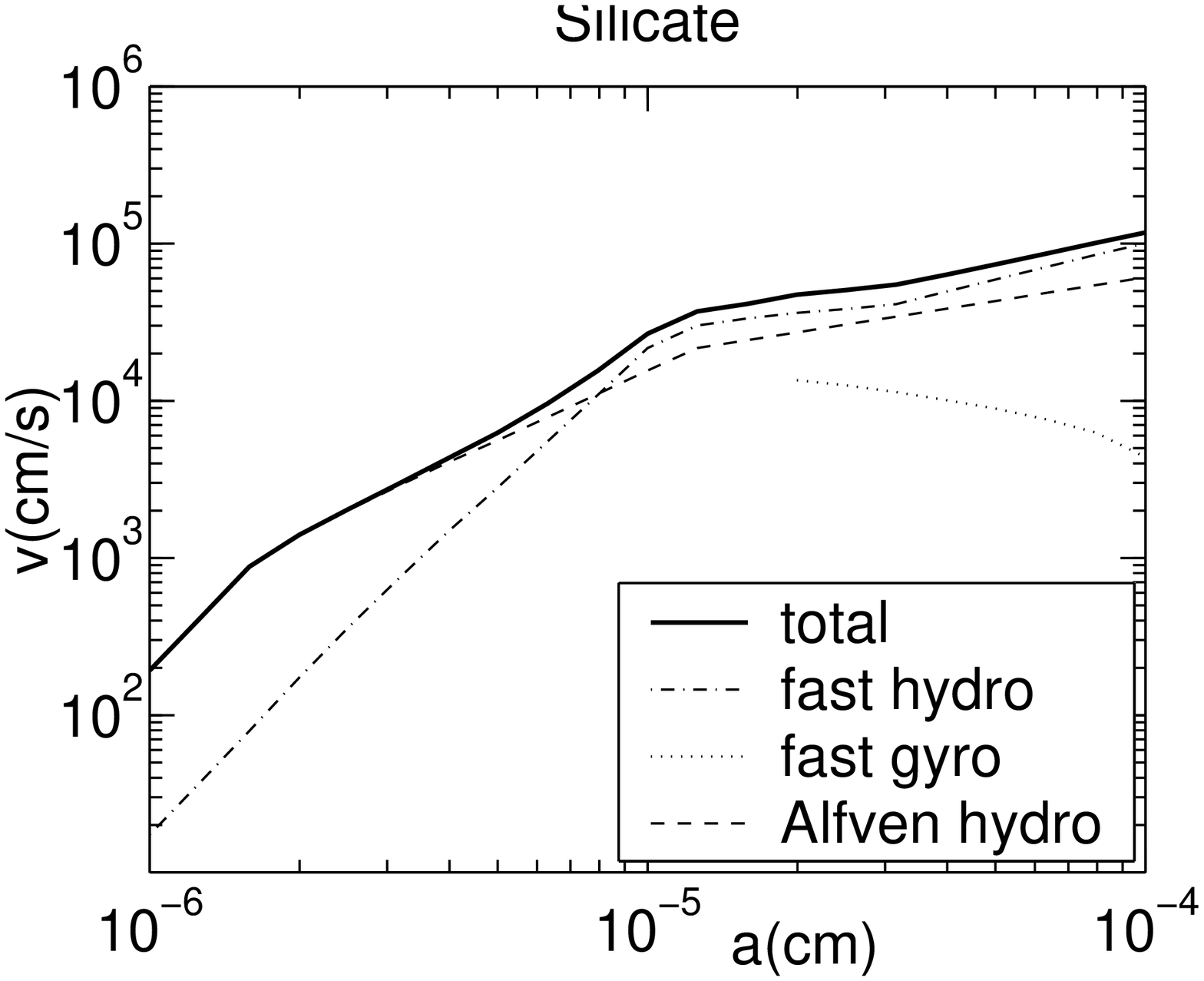} \hfil
 \includegraphics[%
  width=0.49\columnwidth]{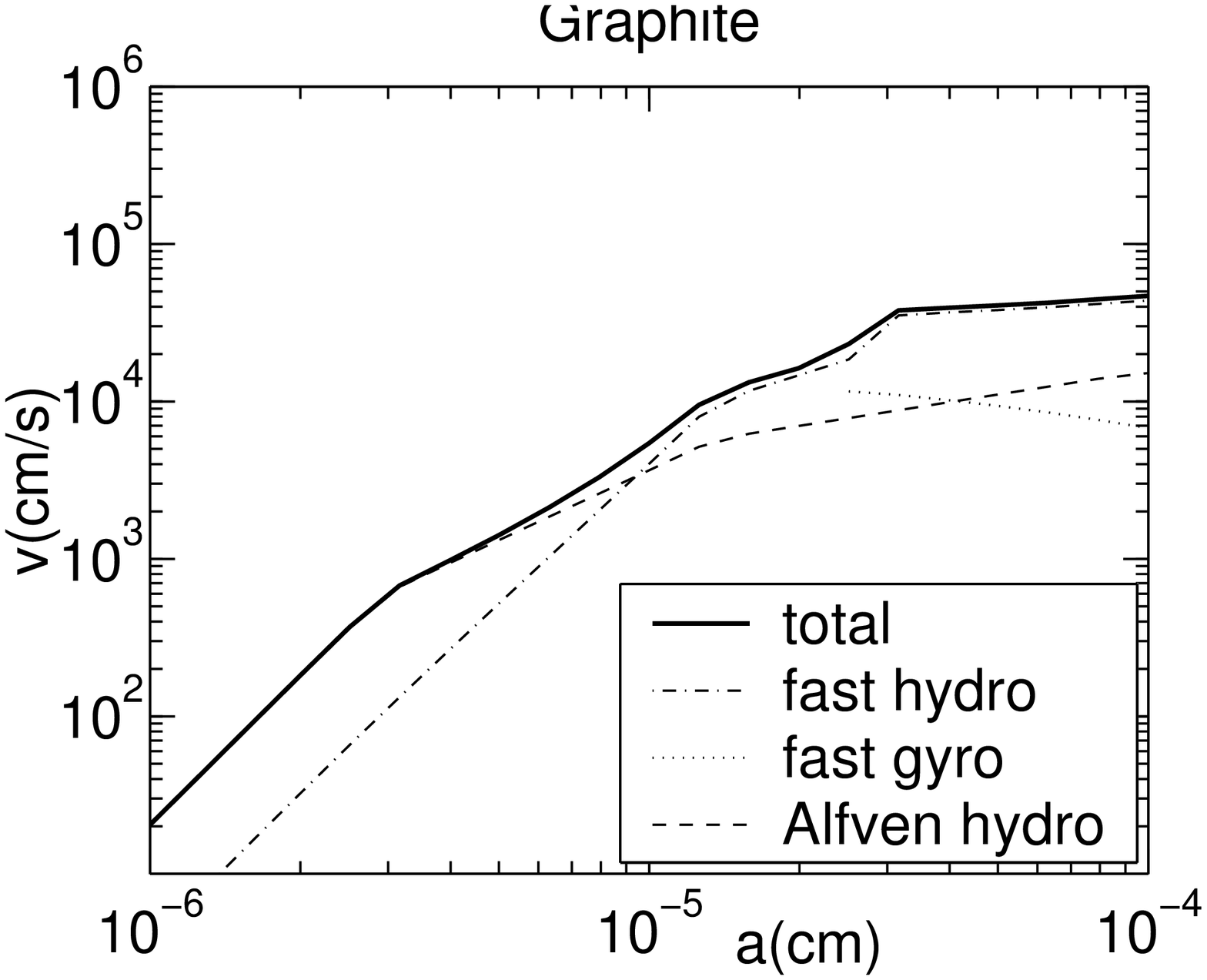}

\caption{Same as Fig.2, but in DC2.}
\end{figure}
It is shown in Fig.~5,6\&7 that the acceleration by gyroresonance in both
MC and DC 
is not as effective as in the lower density media, for two reasons.
First, the low levels of UV and low temperatures result in
reduced grain charge (see Fig.\ 1a).
Secondly, because of the increased density, the drag time $t_d$ is reduced.

The grain velocities in CNM found here are smaller than in an
earlier calculation (see YL03) because we adopt
smaller values for magnetic field 
$B$ and injection velocity $V$.

It should be noted that the strength of magnetic fields in 
the ISM is
still somewhat uncertain and may vary from place to place. We adopted
a particular set of values in above calculations. How would the results
vary as the magnetic field strength varies?
First of all, we know that
the critical condition for acceleration is $k_{res}>k_{d}$: grains
with $k_{res} < k_d$ cannot be accelerated. Thus the cutoff
grain radius $a_c$ varies with the medium, 
$a_c \approx [3 q(a_c) \tau_c B/4\pi\rho]^{1/3}$;
grains with $a<a_c$ are not subject to gyroresonant acceleration.

The magnitude of the velocity is a complex function of
the magnetic field. For illustration,  in Fig.7 we show the grain velocities calculated for magnetic fields a factor of 3 stronger or weaker than the values in Table 1 for the CNM and DC1 environment. Since the hydro
drag by fast modes decreases with the magnetic field, the relative
importance of gyroresonance and hydro drag depends on the magnitude of the magnetic field. In magnetically dominant
regions, gyroresonance is dominant. In weakly magnetized regions, the
frictional drag provides the highest acceleration rate. The injection
scale is another uncertain parameter, but the grain velocity is not
so sensitive to it provided that the injection scale is much larger
than the damping scale.

\begin{figure}
\centering \leavevmode
\includegraphics[%
  width=0.49\columnwidth]{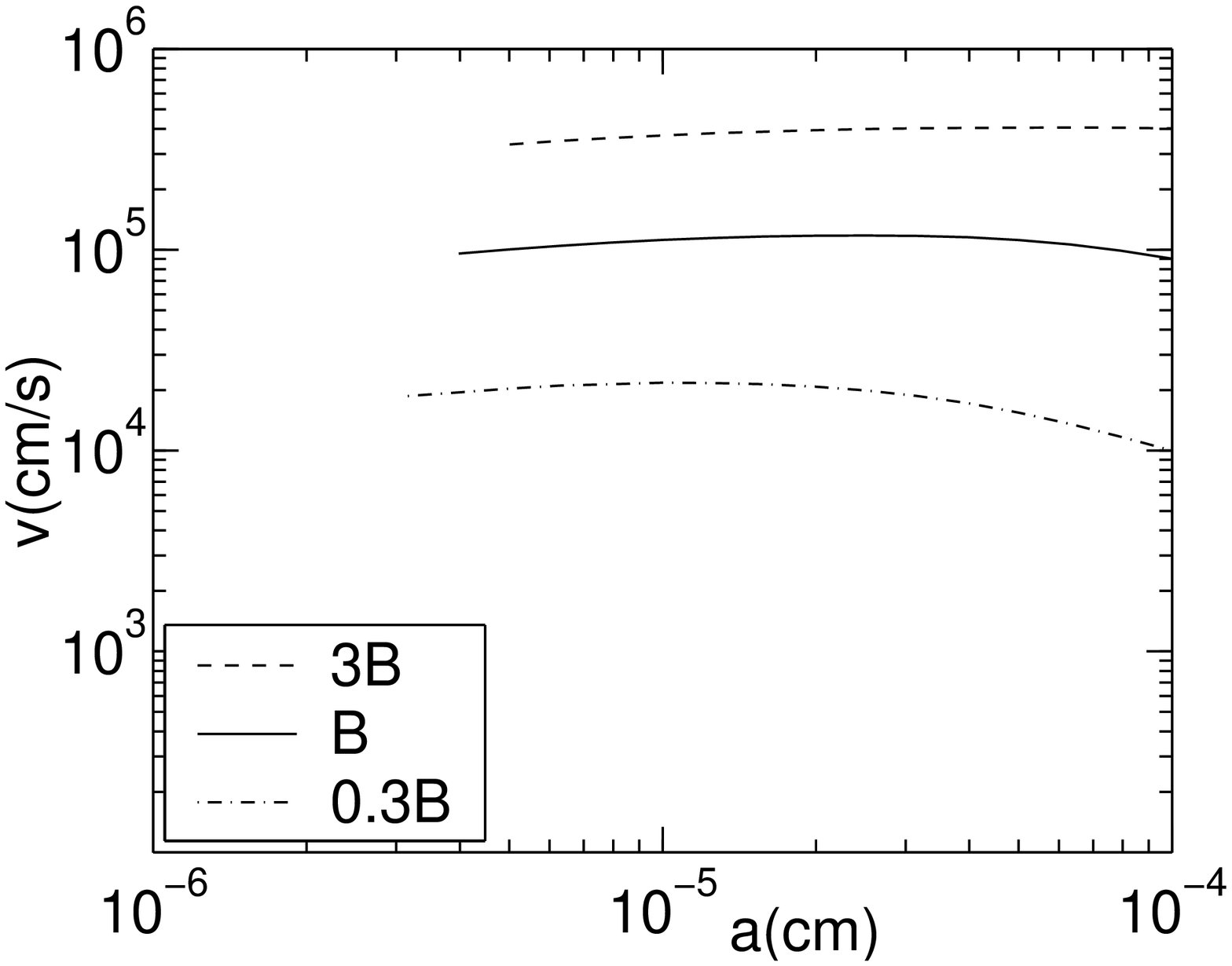} \hfil
 \includegraphics[%
  width=0.49\columnwidth]{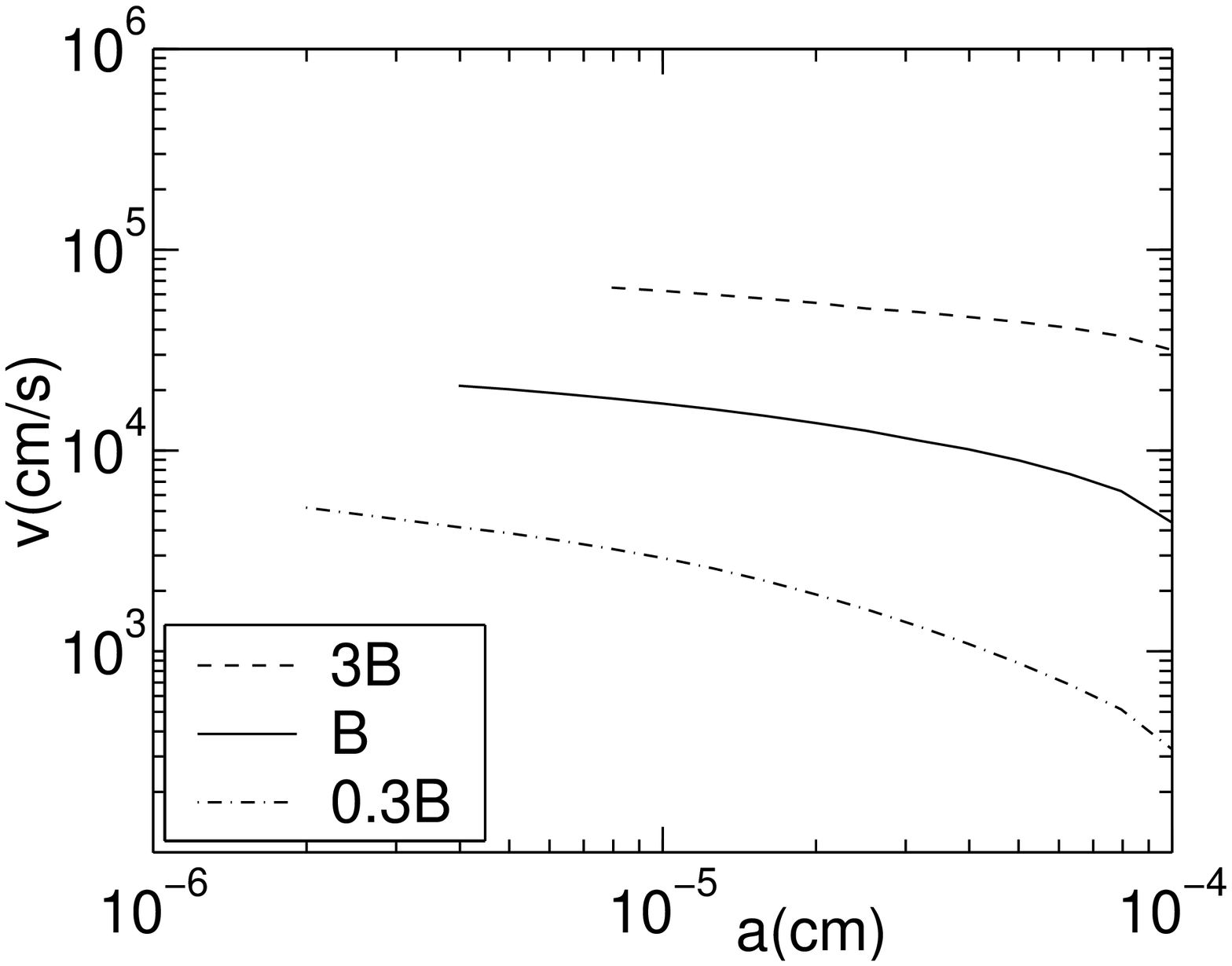}

\caption{Relative velocities gained from gyroresonance as a function of radii
 for different magnetic field strength, \emph{left}: in
CNM, \emph{right}: in DC1. Solid lines are the results for the precedent values of magnetic field. Dashed lines refer to the results with 3 times stronger magnetic field. Dash-dot lines represent the cases with 3 times weaker magnetic field. }
\end{figure}

\section{Discussion}

\emph{Shattering and Coagulation}

With the grain relative velocities known, we can make predictions
for grain shattering and coagulation. For shattering, we adopt the
Jones et al.\ (1996) results, namely, for equal-sized particles
the shattering threshold is
2.7km/s for silicate grains and 1.2km/s for carbonaceous grains. The critical sticking velocity was given by Chokshi et al.\ (1993)
(see also Dominik \& Tielens 1997)%
\footnote{Note a misprint in the exponent of Young's modulus in eq.(10) of
Dominik \& Tielens (1997).}:
$$
v_{cr}= 2.14 F_{\rm stick} \left[\frac{a_1^3+a_2^3}{(a_1+a_2)^3}\right]^{1/2}\frac{\gamma^{5/6}}{E^{1/3}R^{5/6}\rho^{1/2}},
$$
is the maximum relative velocity for coagulation of equal-size spherical 
grains, where $\gamma$ is the surface energy per unit area, $R=a_1a_2/(a_1+a_2)$
is the reduced radius of the grains and $E$ is related to Poisson's ratio $\nu_i$ and Young's modulus $E_i$
by $1/E=[(1-\nu_1)^2/E_1+(1-\nu_2)^2/E_2]$ and we have introduced a factor $F_{stick}\approx 10$ since the
experimental work by Blum (2000) shows that the critical velocity is an
order of magnitude higher than the theoretical estimate of
Chokshi et al.
We use $\gamma$, $E$ and $\nu$ for SiO$_2$ and graphite from Table 1 of Dominik
\& Tielens (1997), and consider collisions between equal-size grains ($a_1\approx a_2$).
Comparing these critical velocities with the velocity curve we obtained
for various media, we can get the corresponding critical size for
each of them (see Table 2). 

\begin{table*}

\begin{tabular}{|c|c|c|c|c|c|c|c|c|c|c|}
\hline 
ISM&
\multicolumn{2}{|c|}{CNM}&
\multicolumn{2}{|c|}{WNM}&
\multicolumn{2}{|c|}{WIM}&
\multicolumn{2}{|c|}{DC1}&
\multicolumn{2}{|c|}{DC2}\tabularnewline
\hline
\hline
Material&
sil.&
  C&
sil.&
 C&
sil.&
 C&
sil.&
C&
sil.&
C\tabularnewline
\hline
Shattering ($\mu$m)&
	NA&
	NA&
  $>0.2$&
 $>0.2$&
$>0.003$&
 $>0.001$&
NA&
NA&
NA&
NA\tabularnewline
\hline
Coagulation ($\mu$m)&
 $<0.01$&
   $<0.02$&
 $<0.02$&
 $<0.04$&
NA&
NA&
$\lesssim0.01$&
$<0.02$&
$<0.02$&
$<0.04$ \tabularnewline
\hline
\end{tabular}
\caption{Size ranges for shattering and coagulation in different medium. NA=not applicable.}
\end{table*}

\emph{Correlation between turbulence and grain sizes}

The grain velocities are strongly dependent on the maximal
velocity of turbulence $V$ at the injection scale, which is highly
uncertain. The critical coagulation and shattering sizes thus
also depend on the amplitude of the turbulence. Variations in the level of turbulence could lead to regional differences in the grain size distribution.

\emph{Elements in cosmic rays}

It has been shown that the
composition of galactic cosmic rays appears to be correlated with elemental volatility (Ellison,
Drury \& Meyer 1997). The more refractory elements are systematically
overabundant relative to the more volatile ones. This suggests that
the material locked in grains must be accelerated more efficiently
than gas-phase ions (Epstein 1980; Ellison, Drury \& Meyer 1997).
The stochastic acceleration of grains, in this case, can act as a
preacceleration mechanism. The ions released from the 
grains in the shock by sputtering or in grain-grain collisions can then
be further accelerated in the shock,
and explain the overabundance of
refractory elements in galactic cosmic rays.

\emph{Heavy Element Depletion and Grain Alignment}

Our results indicate that grains can become supersonic through interaction
with fast modes. 
Grains moving with velocities larger that those of
heavy ions could sweep up
heavy elements which may be advantageous from the point of explaining 
observations (Wakker \& Mathis 2000). Our 
calculations show that while such velocities are readily achievable,
the sign of charging may present a problem for such ``vacuum cleaning''
of the ISM. For instance, silicate grains in MC 
can be accelerated to
 $\gtrsim 4\times10^{4}$~cm/s, which is larger than the thermal speed of heavy ion. Therefore the capture rate for ions by positively charged grains ($\lesssim 2\times 10^{-6}$~cm) would be 
increased. Grains smaller than $2\times 10^{-6}$~cm  will be negatively charged in the MC. For such grains the cross section for Coulomb capture of ions will decrease. If such small grains retain captured ions, this would result
in a decrease of the rate of depletion of metals on grains.  
The actual rates of depletion
on fast moving grains are important and will be identified elsewhere for particular phases of the ISM.

Grains moving supersonically can be aligned mechanically (see a review by Lazarian
2003 and references therein). As pointed out earlier, the scattering
is not efficient for slowly moving grains so that we may ignore the
effect of scattering on the pitch angle. Since
the acceleration of grains increases with the pitch angle of the grain(see
Eq.(\ref{epsilon}) and (\ref{spep})), the supersonic 
grains will tend to have large pitch angles.
As first discussed by Gold (1952), 
gas drag acting on these grains will tend to cause them to spin
with angular
momenta perpendicular to their motion, and therefore tending to be
parallel to the magnetic field direction. 
Dissipational processes will tend to orient the spinning grains with their long
axes perpendicular to their angular momentum, resulting in grain alignment 
with long axes perpendicular to the
magnetic direction.

\emph{Grain Segregation and Turbulent Mixing}

Our results are also relevant to grain segregation. Grains are the
major carrier of heavy elements in the ISM. The issue of grain
segregation may have significant influence on the ISM
metalicity. Subjected to external forcing (WD01, Ciolek and Mouschovias 1996),
grains gain size-dependent velocities with respect to gas. WD01 
considered the forces on dust grains exposed to anisotropic
interstellar radiation fields, including
photoelectric emission, photodesorption, and radiation pressure, and calculated the
drift velocity for grains of different sizes. The velocities they got
for silicate grains in the CNM range from $0.1$cm/s to
$10^{3}$cm/s. Grains can
move along magnetic field lines due to the uncompensated forces,
e.g. due to active sites of H$_2$ formation (see
P79; Lazarian \& Yan 2002b)\footnote{These forces would be mitigated in
molecular clouds, which would induce inflow of dust into molecular
cloud. The latter would affect metalicity of the newborn stars.} Fig.~2a shows that the turbulence produces larger
velocity dispersions\footnote{Our calculation show that for the chosen
set of parameters the effects of systematic thrust are also limited
(see LY02, Lazarian \& Yan 2002b).}. Those velocities are
preferentially perpendicular to magnetic field, but in many cases the
dispersion of velocities parallel to magnetic field will be comparable
to the regular velocities above. This dispersion stems from both the fact
that the transpositions of matter by fast modes are not exactly
perpendicular to magnetic field (see plot in Lazarian \& Yan 2002b)
and due to randomization of directions of grain velocities by
magnetized turbulence (YL03).

More important is that if reconnection in turbulent medium
is fast (see Lazarian \&
Vishniac 1999, Lazarian et al. 2004), 
the mixing of grains over large scales is provided by
turbulent diffusivity $\sim VL/4$.  Usually it was assumed that the
magnetic fields strongly suppress the diffusion of charged species
perpendicular to their directions. However, this assumption is
questionable if we notice that motions perpendicular to the local
magnetic field are hydrodynamic to high order as suggested by Cho,
Lazarian \& Vishniac (2002). Recent work by Cho et al. (2003) found
that the diffusion processes in MHD turbulence are almost as effective
as in the hydrodynamic case if the mean magnetic field is weak or
moderately strong (i.e., $B \le$ the equipartition value)
which would imply that grains can be mixed by the MHD
turbulence. Lazarian \& Yan (2003) therefore concluded that the
segregation of very small and large grains speculated in de
Oliveira-Costa et al. (2002) is unlikely to happen for typical
interstellar conditions.

\section{Summary}

We have calculated the relative motions of dust grains in a magnetized
turbulent fluid. It has been known for decades that 
turbulence can give rise to significant grain-grain velocities.
However, earlier
treatments disregarded the magnetic field and used Kolmogorov turbulence.
Magnetohydrodynamic (MHD) turbulence includes both fluid motions
and magnetic fluctuations. While the fluid motions bring about decoupled
motions to grains, the electromagnetic fluctuations can accelerate
grains through resonant interactions. 

Calculations of grain relative
motion are made for different phases of the ISM with
realistic grain charging, and with turbulence 
that is consistent with the velocity dispersions observed in interstellar
gas. 
We account for the cutoff of
the turbulence from various damping processes. We show that fast modes
dominate grain acceleration, and can drive grains to supersonic velocities. Grains are also scattered by gyroresonance interactions.
The scattering rate is less efficient than acceleration for grains
moving with sub-Alfv\'{e}nic velocities. 

Since the grains are preferentially accelerated with large pitch angles,
the supersonic grains tend to be aligned with long axes perpendicular to
the magnetic field.

Gyroresonant acceleration
is bound to preaccelerate grains that 
will then be further accelerated by shocks.
Grain-grain collisions and sputtering in the shocks 
will inject suprathermal ions which can then undergo further
acceleration in the shock, potentially accounting for the
observed excess of refractory elements in the composition of
galactic cosmic rays (e.g., Epstein 1980; Ellison et al.\ 1997)

\begin{acknowledgments}
We thank Robert Lupton for the SM software package.
AL and HY acknowledge support from NSF grant AST0243156 and
that from the Center for Magnetic Self-Organization in the 
Astrophysical and Laboratory Plasmas.
BTD acknowledges partial support from NSF grant AST-9988126.
\end{acknowledgments}

\appendix

\section{Damping of MHD waves}
Below we summarize the damping processes that we consider in the paper.

\emph{Neutral-ion damping}

In partially ionized medium, a combination of neutral viscosity
and ion-neutral collisional coupling provides damping (see LY02). If
the mean free path for a neutral, $l_{n}$, in a partially ionized
gas with density $n_{tot}=n_{n}+n_{i}$ is much less than the size
of the eddies under consideration, i.e. $l_{n}k\ll1$, the damping
time

\begin{equation}
t_{damp}\sim\nu_{n}^{-1}k^{-2}\sim\left(\frac{n_{tot}}{n_{n}}\right)(l_{n}v_{n})^{-1}k^{-2},\label{tdamp}\end{equation}
 where $\nu_{n}$ is effective viscosity produced by neutrals%
\footnote{The viscosity due to ion-ion collisions is typically small as ion
motions are constrained by the magnetic field. %
}, $v_{n}$ is the thermal velocity of the
neutrals, and
the mean free path of a neutral $l_{n}$ is influenced both by collisions
with neutrals and with ions.
The rate at which neutrals collide with ions is proportional
to the density of ions, while the rate at which neutrals collide with
other neutrals is proportional to the density of neutrals. The momentum transfer rate coefficient for neutral-neutral collisions is 
$\sim1.7\times10^{-10}(T/{\rm K})^{0.3}$~cm$^{3}$ s$^{-1}$ 
(Spitzer 1978), while for neutral-ion collisions
it is $\sim{\langle v_{r}\sigma_{in}\rangle}\approx1.9\times10^{-9}$
cm$^{3}$ s$^{-1}$ (Draine, Roberge \& Dalgarno 1983). Thus collisions
with other neutrals dominate for $n_{i}/n_{n}$ less than $\sim0.09T^{0.3}$.

\emph{Effects of charged grains}

Magnetic perturbations can get decoupled from the fluid motions because 
neutrals are imperfectly coupled to the ions in partially ionized medium\footnote{
We do not discuss here a viscosity-damped
regime of MHD turbulence that takes place in
the partially ionized gas below the scale at which viscosity damps
kinetic motions associated with magnetic field  
(see theory of this regime in Lazarian, Vishniac \& Cho 2004)}. The coupling between ions and neutrals is determined by the ion-neutral collisional rate: 
\begin{equation}
t_{ni}^{-1}=\frac{m_i}{m_n+m_i}n_i \langle v_r\sigma_{in} \rangle
\label{colli}
\end{equation}
where $v_r$ is the ion-neutral relative velocity, $\sigma_{in}$ is the ion-neutral collisional cross section, $m_i$ and $m_n$ are the typical ion and neutral masses, $n_i$ is the ion number density. When the collisional time $t_{ni}$ is equal to the wave period, neutrals are decoupled from magnetic field, and
turbulence becomes hydrodynamic. In molecular clouds, grains can take substantial portion of the total charge.  The contribution of charged grains to coupling neutrals 
to magnetic fields depends on grain size spectrum. The ratio of the ion-neutral collisional rate to the grain-neutral collisional rate is (Nishi \& Nakano 1991, Elmegreen \& Fiebig 1993):

\begin{eqnarray}
\frac{t_{ni}^{-1}}{
\langle n_g\sigma_g \rangle v_n}\simeq \cases{0.25n_4^{-1/2};& (MRN)\cr
 3.3n_4^{-1/2};& (MW) \cr},
\end{eqnarray}
where MRN refers to the grain size distribution proposed by Mathis et al. (1977) and MW stands for the distribution suggested by Mathis \& Whiffen (1989). From this expression, we can see that ions are always the dominant contribution for the coupling in MC. In DC, the situation will depend on grain size distribution. DC is a denser region where observations favor MW distribution (Elmegreen \& Fiebig 1993). Thus presumably the contribution from grains in DC is also subdominant and we neglect it in the main text. 

\emph{Collisionless damping}

The nature of collisionless damping is closely related to the radiation
of charged particles in magnetic field. Since the charged particles
can emit plasma waves through acceleration (cyclotron radiation) and
Cherenkov effect, they also absorb the radiation under the same condition
(Ginzburg 1961). The damping rate $\gamma_{d}=\tau_{d}^{-1}$
of the fast modes of frequency $\omega$ for $\beta\ll1$ and $\theta\sim1$
(Ginzburg 1961) is

\begin{eqnarray}
\Gamma_{d} & = & \frac{\sqrt{\pi\beta}}{4}\omega\frac{\sin^{2}\theta}{\cos\theta}\times\left[
\sqrt{\frac{m_{e}}{\mH}}\exp(-\frac{m_{e}}{\mH\beta\cos^{2}\theta}) +5\exp(-\frac{1}{\beta\cos^{2}\theta})
\right]
,\end{eqnarray}
where $m_{e}$ is the electron mass. The exact expression for the
damping of fast waves at small $\theta$ was obtained in Stepanov%
\footnote{We corrected a typo in the corresponding expression.%
} (1958)

\[
\Gamma_{d}=\frac{\sqrt{\pi\beta}}{4}\omega\theta^{2}\times\left(1+\frac{\theta^{2}}{\sqrt{\theta^{4}+4\Omega_{i}^{2}/\omega^{2}}}\right)\sqrt{\frac{m_{e}}{\mH}}\exp(-\frac{m_{e}}{\mH\beta\cos^{2}\theta}).\]
 When $\beta\gg1$ (see Foote \& Kulsrud 1979), 
\begin{eqnarray}
\Gamma_{L}=\cases{2\omega^{2}/\Omega_{i}  &  for $k<\Omega_{i}/\beta V_{A}$\cr
2\Omega_{i}/\beta  &  for $k>\Omega_{i}/\beta V_{A}$\cr}
\label{hbcoll}
\end{eqnarray}
where $\Omega_{i}$ is the ion gyrofrequency.

\emph{Ion viscosity}

In a strong magnetic field ($\Omega_{i}\tau_{i}\gg1$) the transport
of transverse momentum is prohibited by the magnetic field (along $\hat z$). Thus transverse viscosity $\eta_{\perp}$ is much smaller than longitudinal viscosity
$\eta_{0}$, $\eta_{\perp}\sim\eta_{0}/(\Omega_{i}\tau_{i})^{2}$.
Following Braginskii (1965), we can find the damping rate is(see Yan \& Lazarian 2004):

\begin{eqnarray}
\Gamma_{ion}=\cases{	k_{\perp}^{2}\eta_{0}/6\rho_{i}   & for $\beta\ll 1$\cr
					k^{2}\eta_{0}(1-3\cos^{2}\theta)^{2}/6\rho_{i} &  for $\beta\gg 1$\cr}
\label{viscous}
\end{eqnarray}

For more discussion, see Yan \& Lazarian (2004).

\section{Fokker-Planck coefficients}

In quasi-linear theory (QLT), the effect of MHD waves is studied by
calculating the first order corrections to the particle orbit in the
uniform magnetic field, and the ensemble-averaging over the statistical
properties of the MHD waves (Jokipii 1966, Schlickeiser \& Miller
1998). Obtained by applying the QLT to the collisionless Boltzmann-Vlasov
equation, the Fokker-Planck equation is generally used to describe
the evolvement of the gyrophase-averaged particle distribution,
\[
\frac{\partial f}{\partial t}=\frac{\partial}{\partial\mu}\left(D_{\mu\mu}\frac{\partial f}{\partial\mu}+D_{\mu p}\frac{\partial f}{\partial p}\right)+\frac{1}{p^{2}}\frac{\partial}{\partial p}\left[p^{2}\left(D_{\mu p}\frac{\partial f}{\partial\mu}+D_{pp}\frac{\partial f}{\partial p}\right)\right],\]
 where $p$ is the particle momentum. The
Fokker-Planck coefficients $D_{\mu\mu},D_{\mu p},D_{pp}$ are the
fundamental physical parameter for measuring the stochastic interactions,

\begin{eqnarray}
\left(\begin{array}{c}
D_{\mu\mu}\\
D_{\mu p}\\
D_{pp}\end{array}\right) & = & {\frac{\pi\Omega^{2}(1-\mu^{2})}{2}}\int_{\bf k_{min}}^{\bf k_{max}}dk^3\frac{\tau_{k}^{-1}}{\tau_{k}^{-2}+(\omega-k_{\parallel}v\mu-\Omega)^{2}}\left(\begin{array}{c}
\left(1+\frac{\mu V_{ph}}{v\zeta}\right)^{2}\\
\left(1+\frac{\mu V_{ph}}{v\zeta}\right)mV_{A}\\
m^{2}V_{A}^{2}\end{array}\right)\label{gyroapp}\\
 &  & \left\{ (J_{2}^{2}({\frac{k_{\perp}v_{\perp}}{\Omega}})+J_{0}^{2}({\frac{k_{\perp}v_{\perp}}{\Omega}}))\left[\begin{array}{c}
M_{{\mathcal{RR}}}({\mathbf{k}})+M_{{\mathcal{LL}}}({\mathbf{k}})\\
-C_{{\mathcal{RR}}}({\mathbf{k}})-C_{{\mathcal{LL}}}({\mathbf{k}})\\
K_{{\mathcal{RR}}}({\mathbf{k}})+K_{{\mathcal{LL}}}({\mathbf{k}})\end{array}\right]\right.\nonumber \\
 & - & 2J_{2}({\frac{k_{\perp}v_{\perp}}{\Omega}})J_{0}({\frac{k_{\perp}v_{\perp}}{\Omega}})\left.\left[e^{i2\phi}\left[\begin{array}{c}
M_{{\mathcal{RL}}}({\mathbf{k}})\\
-C_{{\mathcal{RL}}}({\mathbf{k}})\\
K_{{\mathcal{RL}}}({\mathbf{k}})\end{array}\right]+e^{-i2\phi}\left[\begin{array}{c}
M_{{\mathcal{LR}}}({\mathbf{k}})\\
-C_{{\mathcal{LR}}}({\mathbf{k}})\\
K_{{\mathcal{LR}}}({\mathbf{k}})\end{array}\right]\right]\right\}, \nonumber \end{eqnarray}
where $|{\bf k_{min}}|=k_{\min}=L^{-1}$, $|{\bf k_{max}}|=k_{\max}$ corresponds to the dissipation scale, $\mathcal{R,L}$ refer to left- and right-circularly polarized modes, and $\phi=\tan^{-1}k_x/k_y$.

The correlation tensors are defined as following:
\begin{equation}
\begin{array}{c}
<B_{\alpha}(\mathbf{k},t)B_{\beta}^{*}(\mathbf{k'},t+\tau)>/B_{0}^{2}=\delta(\mathbf{k}-\mathbf{k'})M_{\alpha\beta}(\mathbf{k})e^{-\tau/\tau_{k}}\\
<v_{\alpha}(\mathbf{k},t)B_{\beta}^{*}(\mathbf{k'},t+\tau)>/V_{A}B_{0}=\delta(\mathbf{k}-\mathbf{k'})C_{\alpha\beta}(\mathbf{k})e^{-\tau/\tau_{k}}\\
<v_{\alpha}(\mathbf{k},t)v_{\beta}^{*}(\mathbf{k'},t+\tau)>/V_{A}^{2}=\delta(\mathbf{k}-\mathbf{k'})K_{\alpha\beta}(\mathbf{k})e^{-\tau/\tau_{k}},\end{array}\end{equation}
 where $B_{\alpha,\beta}$, $v_{\alpha,\beta}$ are respectively the
magnetic and velocity perturbation associated with the turbulence,$\tau_{k}$
is the nonlinear decorrelation time and essentially the cascading
time of the turbulence. For the balanced cascade we consider (see
discussion of our imbalanced cascade in CLV02), i.e., equal intensity
of forward and backward waves, $C_{ij}(\mathbf{k})=0$.

The magnetic correlation tensor for Alfv\'{e}nic turbulence is (CLV02),

\begin{eqnarray}
\left[\begin{array}{c}
M_{ij}({\mathbf{k}})\\
K_{ij}({\mathbf{k}})\end{array}\right]&=&\frac{L^{-1/3}}{12\pi}I_{ij}k_{\perp}^{-10/3}\exp(-L^{1/3}|k_{\parallel}|/k_{\perp}^{2/3}),\nonumber\\
\tau_{k}&=&(L/V_{A})(k_{\perp}L)^{-2/3}\sim k_{\parallel}V_{A}\label{anisotropic}\end{eqnarray}
where $I_{ij}=\{\delta_{ij}-k_{i}k_{j}/k^{2}\}$ is a 2D tensor in
$x-y$ plane which is perpendicular to the magnetic field, $L$ is
the injection scale, $V$ is the velocity at the injection scale.
Slow modes are passive and similar to Alfv\'{e}n modes. The normalization
constant is obtained by assuming equipartition $\epsilon_{k}=\int dk^{3}\sum_{i=1}^{3}M_{ii}B_{0}^{2}/8\pi\sim B_{0}^{2}/8\pi$.
The normalization for the following tensors below are obtained in
the same way.

According to CL02, fast modes are isotropic and have one dimensional
energy spectrum $E(k)\propto k^{-3/2}$. In low $\beta$ medium, the
corresponding correlation is (YL03)

\begin{eqnarray}
\left[\begin{array}{c}
M_{ij}({\mathbf{k}})\\
K_{ij}({\mathbf{k}})\end{array}\right]={\frac{L^{-1/2}}{8\pi}}H_{ij}k^{-7/2}\left[\begin{array}{c}
\cos^{2}\theta\\
1\end{array}\right],~~
\tau_{k}=(k/L)^{-1/2}\times V_{A}/V^{2},
\label{fast}\end{eqnarray}
where $\theta$ is the angle between $\mathbf{k}$ and $\mathbf{B}$,
$H_{ij}=k_{i}k_{j}/k_{\perp}^{2}$ is also a 2D tensor in $x-y$ plane.
The factor $\cos^{2}\theta$ represents the projection as magnetic
perturbation is perpendicular to $\mathbf{k}$. This tensor is different
from that in Schlickeiser \& Miller (1998). For isotropic turbulence,
the tensor of the form $\propto E_{k}(\delta_{ij}-k_{i}k_{j}/k^{2})$
was obtained to satisfy the divergence free condition $\mathbf{k}\cdot\delta\mathbf{B}=0$
(see Schlickeiser 2002). Nevertheless, the fact that $\delta\mathbf{B}$
in fast modes is in the $\mathbf{k}$-$\mathbf{B}$ plane places another
constraint on the tensor so that the term $\delta_{ij}$ doesn't exist. 

In high $\beta$ medium, fast modes in this regime are essentially
sound waves compressing the magnetic field (Goldreich \& Sridhar 1995,
Lithwick \& Goldreich 2001, CL03). The compression of magnetic field
depends on plasma $\beta$. The corresponding x-y components of the
tensors are 

\begin{eqnarray}
\left[\begin{array}{c}
M_{ij}({\mathbf{k}})\\
K_{ij}({\mathbf{k}})\end{array}\right]={\frac{L^{-1/2}}{2\pi}}\sin^{2}\theta H_{ij}k^{-7/2}\left[\begin{array}{c}
\cos^{2}\theta/\beta\\
1/\end{array}\right],~~
\tau_{k}=(k\times k_{min})^{-1/2}\times C_{s}/V^{2},\label{hbeta}
\end{eqnarray}
where $C_{s}$ is the sound speed. The velocity perturbations in high
$\beta$ medium are longitudinal, i.e., along $\mathbf{k}$, thus we have
the factor $\sin^{2}\theta$ and also a factor $V_{A}^{2}/C_{s}^{2}=2/\beta$
from the magnetic frozen condition $\omega\delta\mathbf{B}\sim\mathbf{k}\times(\mathbf{v}_{k}\times\mathbf{B})$.
We use these statistics to calculate
grain acceleration arising from MHD turbulence.

The spherical components of the correlation tensors are obtained below.

For Alfv\'{e}n modes, their tensors are proportional to

\begin{center}
\begin{equation}
I_{ij}=\left(\begin{array}{cc}
\sin^{2}\phi & -\cos\phi\sin\phi\\
-\cos\phi\sin\phi & \cos^{2}\phi\end{array}\right).
\end{equation}
\end{center}
Thus we get

\begin{eqnarray}
I_{{\mathcal{RR}}}=I_{{\mathcal{LL}}} = {\frac{(I_{x}-iI_{y})}{\sqrt{2}}}{\frac{(I_{x}^{*}+iI_{y}^{*})}{\sqrt{2}}}={\frac{1}{2}}(I_{xx}+I_{yy})={\frac{1}{2}},\nonumber
\label{ARRLL}\end{eqnarray}
and

\begin{eqnarray}
e^{i2\phi}I_{{\mathcal{RL}}}+e^{-i2\phi}I_{{\mathcal{LR}}}={\frac{(I_{x}-iI_{y})^{2}}{2}}\times e^{i2\phi}+{\frac{(I_{x}+iI_{y})^{2}}{2}}\times e^{-i2\phi}=(I_{xx}-I_{yy})\cos2\phi+(I_{xy}+I_{yx})\sin2\phi=-1.\nonumber
\label{ARL}
\end{eqnarray}

For fast modes, their tensors have such a component

\begin{eqnarray}
H_{ij}=Ak^{-3.5}\left(\begin{array}{cc}
\cos^{2}\phi & \cos\phi\sin\phi\\
\cos\phi\sin\phi & \sin^{2}\phi\end{array}\right).\end{eqnarray}
Thus we have

\begin{eqnarray}
H_{{\mathcal{RR}}}=H_{{\mathcal{LL}}} & = & {\frac{1}{2}}(H_{xx}+H_{yy})=\frac{1}{2},\nonumber
\label{FRRLL}\end{eqnarray}
and

\begin{eqnarray}
e^{i2\phi}H_{{\mathcal{RL}}}+e^{-i2\phi}H_{{\mathcal{LR}}} & =&(H_{xx}-H_{yy})\cos2\phi+(H_{xy}+H_{yx})\sin 2\phi=1.\nonumber
\label{FRL}
\end{eqnarray}

\section{Drag due to the dipole moment of grain}
The plasma drag to the rotational motion for stationary grains has
been considered earlier (Anderson \& Watson 1993, Draine \& Lazarian
1998, henceforth DL98). Similarly, for a grain with electric dipole moment $\mu$,
there also exist forces between the grain and nearby ions. Consider the effects in the comoving
frame of grain. In this frame ions move at speed $\mathbf{v}$ with
impact parameter \emph{b}. To simplify the problem, we define a stopping
cross section $\sigma_{s}=\pi b_{0}^{2}$, where $b_{0}$ is defined
by $Z_{i}e\mu/b_{0}^{2}=m_{i}v^{2}.$ For impact parameters $b<\max(a,b_0)$,
the interaction is strong. 
We will assume 
\begin{equation}
a < b_0 = \left(\frac{Z_i e \mu}{m_i v^2}\right)^{1/2} =
5.4\times10^{-7}\cm~ Z_i^{1/2} \mu_1 \left(\frac{\mH}{m_i}\right)^{1/2}
\frac{{\rm km~s}^{-1}}{v}
\end{equation}
where $\mu_1 \equiv \mu/10~{\rm debye}$.
\footnote{In DL98 it was 
%assumed that $\mu$ goes in a random walk fashion which results in $\mu^2\sim a^3$ for grains which dipole moment is dominated by intrinsic dipole moments of polarized bonds. In this case, this assumption is not particularly restrictive, especially for relatively low temperatures.}
estimated that $\mu_1\approx 0.93 (a/10^{-7}\cm)^{3/2}$.}
%---
If we assume the ions to be scattered isotropically, then the
drag force due to strong scattering events is
\begin{equation}
F_s \approx n_i Z_i e \mu \frac{4\sqrt{\pi}}{3} 
\frac{v_{gr}}{\max\left[v_m, (4/3\sqrt{\pi})v_{gr}\right]}
\end{equation}
where $v_m=(2kT/m)^{1/2}$ is the most probable speed of ions at temperature $T$. For ions with impact parameter $>b_{0}$, we assume their trajectories are barely
changed during the collisions with the grain.
Define the direction
of $\mathbf{v}$ as the polar axis $\hat{e}_z$,
 let pericenter be at $(b,0,0)$, and let \emph{t} be the time from pericenter.
The force on the ion from the dipole moment is
\begin{equation}
\textrm{F}_{di}=\frac{Z_{i}e}{(b^{2}+v^{2}t^{2})^{2.5}}(\hat{e}_{x}\mu_{x}(2b^{2}+3vtb-v^{2}t^{2})+\hat{e}_{z}\mu_{z}(2v^{2}t^{2}+3vtb-b^{2})-\hat{e}_{y}\mu_{y}(b^{2}+v^{2}t^{2})).
\end{equation}
Integrated over time from $-\infty$ to $\infty$, this expression
yields the total momentum delivered to the grain:
\begin{equation}
\triangle\mathbf{p}=\frac{2Z_{i}e}{b^{2}v}
(-\hat{e}_{x}\mu_{x}+\hat{e}_{y}\mu_{y}).\label{dpperp}
\end{equation}
$\mathbf{p}$ increases in a random walk fashion, therefore the impulses
of individual collisions should be added in quadrature. If we now
average over random orientation of $\mu$ and then integrate over
impact parameters and thermal distribution of ion velocities, we find
\begin{equation}
\frac{dp^{2}}{dt}=n_{i} \int_{0}^{\infty}dv4\pi v^{2}f_i(v)v\int_{b_{0}}^{b_{2}}2\pi bdb\frac{2}{3}\left(\frac{2\mu Z_{i}e}{b^2v}\right)^{2}
=\frac{16\sqrt{\pi}}{3} n_i Z_i e \mu m_iv_m~~~,
\end{equation}
where for the upper cutoff we take the Debye length
$b_2 = (kT/4\pi n_e e^2)^{1/2} \gg b_0$. Using the fluctuation-dissipation theorem, we then can get
the damping force for subsonic grains,
\begin{equation}
F = F_s + \frac{v_{gr}}{6kT} \frac{dp^2}{dt} =  
\frac{28\sqrt{\pi}}{9} n_i Z_i e\mu
\frac{v_{gr}}{v_m}~~~.
\end{equation}

If the grains becomes supersonic, the fluctuation-dissipation theorem
is no longer applicable. Since the interaction is elastic, the loss
of the momentum in the direction parallel to the moving direction
can be obtained by assuming energy conservation: $p_{z}^{2}+p_{\perp}^{2}=const$. The dipole interaction is weak interaction so that $\triangle p/p\ll1$
during one encounter. Thus we can estimate the momentum loss in the
$z$ direction as $\Delta p_{z}=\triangle p_{\perp}^{2}/2p$, where
$\Delta p_{\perp}$ is given by Eq.(\ref{dpperp}). Then integrating over impact parameter,
we get the damping force
\begin{equation}
F=\pi n_{i}Z_{i}e\mu + \frac{n_{i}}{m_i}\int_{b_{0}}^{b_{2}}
2\pi bdb ~\frac{4}{3}\left(\frac{Z_{i}e\mu}{b^2}\right)^{2}\frac{1}{ v_{gr}^2}
= \frac{7\pi}{3}n_i Z_i e \mu~~~.
\end{equation}

To determine the importance of this force, we compare it
with other drag forces. Using the dipole moment 
estimated by DL98,
we find the dipole drag is smaller than collisional drag.
However, for very small neutral grains with a dipole moment in ionized gas,
dipole
plasma drag may play a more important role. 

We note that the estimate by DL98 of rotational excitation
by ``plasma drag'' acting on the grain dipole moment overestimated the
transfer of angular momentum by using the weak interaction approximation for
all trajectories with $b>a$.
Assuming random scattering, the mean square angular momentum transfer from
strong scattering events will be $\sim 2(m_i v b)^2$, and thus the
contribution of impact parameters $a < b < b_0$ to $dL^2/dt$ is

\begin{eqnarray}
\frac{dL^2}{dt} 
&\approx&
n_i\int_0^\infty dv 4\pi v^2 f_i(v) v \int_a^{b_0} 2\pi b db~ 2(mvb)^2
= 4\pi^2 n_i m^2 \int_0^\infty dv v^5 f_i(v) (b_0^4-a^4)
\nonumber\\
&=&
2\sqrt{\pi} n_i \frac{Z_i^2 e^2 \mu^2}{v_m} 
\left[
1-\frac{2 m^2 v_m^4 a^4}{Z^2e^2\mu^2}
\right]
\approx5.71\times10^{-4}\hbar^2{~\rm s}^{-1} \left(\frac{n_i}{\cm^{-3}}\right)
\left(\frac{m_i}{\mH}\right)^{1/2} T_2^{-1/2} Z_i^2 \mu_1^2
\label{scatter}
\end{eqnarray}
%---
where $T_2\equiv T/100$K, $a_{-7}\equiv a/10^{-7}$cm. 
For comparison, for impact parameters $a < b < b_0$ DL98
found 
\begin{equation}
\frac{dL^2}{dt}=
\frac{32\sqrt{\pi}}{3}n_i
\frac{Z_i^2 e^2\mu^2}{v_m}
\ln\left(\frac{b_{0}}{a}\right)
=3.05\times10^{-3}\hbar^2 {~\rm s}^{-1} 
\left(
\frac{n_i}{\cm^{-3}}
\right)
\left(
\frac{m_i}{\mH}
\right)^{1/2}
 T_2^{-1/2}
Z_i^2\mu_{1}^2
\ln\left[
\frac{4.07}{a_{-7}^{1/4}T_2^{1/2}}
\right].
\label{DL}
\end{equation}

This part of contribution is comparable with the total if 
$a_{-7}\lesssim 1$ and $T_2\lesssim 1$ 
or  if the angle between the dipole moment and the rotation velocity 
is close to $90^\circ$ (see eq.(B35) in DL98). 
For instance, at $T=100$K, for grains with $a_{-7}=0.5$, 
the part given by eq.(\ref{DL}) is $\sim 35\%$ of the total 
for $\cos^2\Psi=1/3$.  
In such cases, the correction owing to the strong scattering 
as given in Eq.(\ref{scatter}) should be taken into account. 

\section{ Angle between $\textbf{B}$ and ${\textbf{v}}$}

A fundamental question arises from the fact that
in MHD turbulence
wave vectors are not aligned along magnetic field lines, as is
the case for pure Alfv\'{e}nic waves. We need to analyze the
relative position of three vectors: magnetic field vector $\textbf{B}$, wave
vector $\textbf{k}$, and the displacement velocity vector $\textbf{v}$.
In what follows, we shall study how the angle $\gamma$ between
${\textbf{v}}$ and $\textbf{B}$ changes with plasma $\beta$ .

\begin{figure}
\centering \leavevmode
\includegraphics[%
  width=0.33\textwidth,
  height=0.25\textheight]{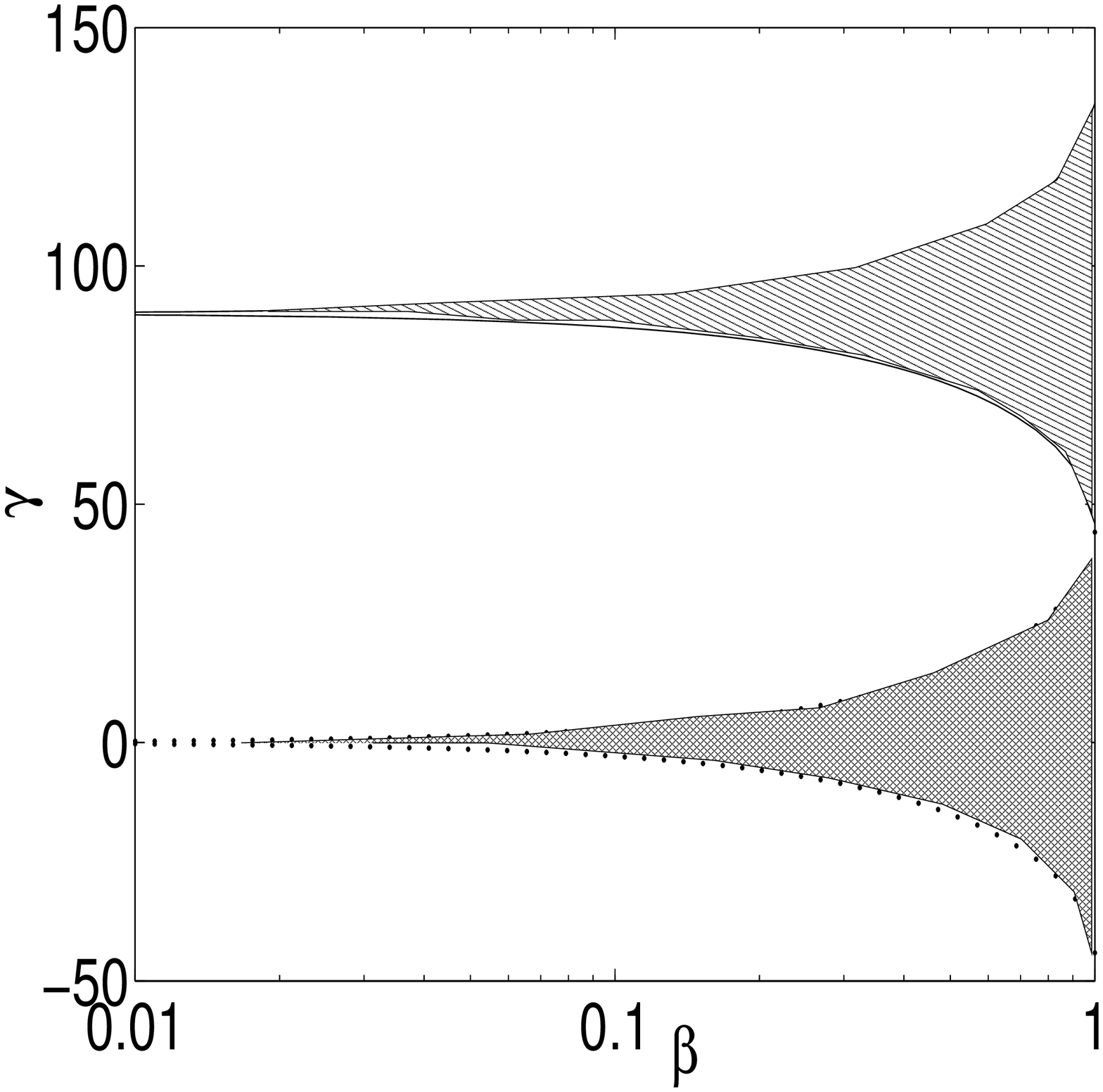}
\caption{The range of angles between \protect\protect$\textbf{B}$ and ${\textbf{v}}$.
The dashed area refers to the fast modes, the meshed area represents the range where slow modes fall in. The ranges are produced when the angle between ${\textbf{k}}$ and ${\textbf{B}}$ changes from $0$
to $\pi$.\label{fm2} }
\end{figure}

It is shown in Alfv\'{e}n \& F\"{a}lthmmar (1963) that the angle
$\Psi$ between ${\textbf{v}}$ and $\textbf{k}$ can be expressed
as follows: \begin{equation}
\tan\Psi=\frac{\sin\theta\cos\theta}{\cos^{2}\theta-v_{p}^{2}/V_{A}^{2}},\label{al1}\end{equation}
where $\theta$ is the angle between ${\bf k}$ and ${\bf B}$,
and the phase velocity $v_{p}$, is related to the Alfv\'{e}nic
velocity $V_{A}$ and the sound velocity $C_{s}$ through the dispersion
relation \begin{equation}
v_{p}^{4}-(V_{A}^{2}+C_{s}^{2})v_{p}^{2}+C_{s}^{2}V_{A}^{2}\cos^{2}\theta=0.\label{al2}\end{equation}
 Solving this equation for $\epsilon=v_{f}^{2}/v_{A}^{2}$,
\begin{equation}
\epsilon(\beta)=\frac{1}{2}\left(1+\beta/2\pm\sqrt{(1-\beta/2)^{2}+2\beta\sin^{2}\theta}\right),\label{al3}\end{equation}
 where '$+$' gives the result for fast mode and '$-$' represents
slow mode. Thus the angle $\gamma$ can be calculated as \begin{equation}
\gamma=\theta-\arctan\frac{\sin\theta\cos\theta}{\cos^{2}\theta-\epsilon(\xi)}\label{al5}\end{equation}
 and the corresponding plot is shown in Fig.~\ref{fm2}. It is evident
that for low $\beta$ plasma the velocity $v$ of the fast mode is
directed nearly perpendicular to $\textbf{B}$ whatever the direction
of $\textbf{k}$, while the velocity of the slow mode is nearly parallel
to the magnetic field. So the parallel motions we got from slow mode
are essentially correct, while the perpendicular motions are also
subjected to fast mode. A more general discussion of the issue is
given in CLV02.

\begin{thebibliography}{10}
\bibitem{dummy}Alfv\'{e}n, H. \& F$\ddot{a}$lthmmar, C.G. 1963, \emph{Cosmical
Electrodynamics}, Oxford, Clarendon Press 
\bibitem{}Anderson, N., \& Watson, W. D. 1993, A \& A, 270, 477
\bibitem{key-24}Armstrong, J. W., Rickett, B. J. \& Spangler, S. R. 1995, ApJ, 443,
209
\bibitem{dummy}Arons, J. \& Max, C.E. 1975, ApJ, 196, L77 
\bibitem{dummy}Biermann, L., \& Harwit, M. 1980, ApJ, 241, L105 
\bibitem{dummy}Black, J.H., \& van Dishoeck, E.F. 1991, ApJ, 369, L9
\bibitem{dummy}Blum, J. 2000, Space Sci. Rev., 92, 265B 
\bibitem{key-52}Boldyrev, S., Nordlund, A. \& Padoan, P. 2002, ApJ, 573, 678
\bibitem{key-19}Braginskii, S.I. 1965, Rev. Plasma Phys. 1, 205 
\bibitem{dummy}Cho, J. \& Lazarian, A. 2002, Phys. Rev. Lett., 88, 245001 (CL02) 
\bibitem{key-28}Cho, J. \& Lazarian, A. 2003a, to appear in Acoustic emission and
scattering by turbulent flows, ed. M. Rast (Springer LNP), astro-ph/0301462
\bibitem{key-29}Cho, J. \& Lazarian, A. 2003b, Rev. Mex. A\&A, Vol. 15, pp. 293 (CL03) 
\bibitem{key-30}Cho, J. \& Lazarian, A. 2003c, MNRAS, 345, 325
\bibitem{key-34}Cho, J, Lazarian, A., Honein, A., Knaepen B., Kassinos, S. \& Moin
P. 2003, ApJ, 589, L77
\bibitem{key-21}Cho, J., Lazarian, A. \& Vishniac, E.T. 2002a, in {}``Turbulence
and Magnetic field in Astrophys'', eds. T. Passot \& E. Falgorone
(Springer LNP), p.56
\bibitem{dummy}Cho, J., Lazarian, A. \& Vishniac, E.T. 2002b, ApJ, 564, 291 (CLV02)
\bibitem{key-54}Cho, J., Lazarian, A. \& Vishniac, E.T. 2002c, ApJL, 566, 49L
\bibitem{dummy}Chokshi, A., Tielens, A.G.G.A. \& Hollenbach, D. 1993, ApJ, 407, 806 
\bibitem{}Ciolek and Mouschovias 1996, ApJ, 468, 749
\bibitem{}Crutcher, 1999, ApJ, 520, 706
\bibitem{dummy}de Oliveira-Costa, A., Tegmark, M., Devies, R.D., Gutierrez, C.M.,
Mark J., Haffner, L.M., Jones, A.W., Lasenby, A.N., Rebolo, R., Reynolds,
R.J., \& Tufte, S.L., Watson, R.A. 2002, ApJ, 567, 363 
\bibitem{dummy}Dolginov, A. Z. \& Mitrofanov, I. G. 1975, Astronomicheskii Zhurnal,
vol. 52, Nov.-Dec. 1975, p. 1268-1278. In Russian. 
\bibitem{dummy}Dominik, C. \& Tielens, A.G.G.A. 1997, ApJ, 480, 647 
\bibitem{dummy}Draine, B.T. 1985, in Protostars and Planets II, ed. D.C. Black \&
M.S. Matthews (Tucson: Univ. Arizona Press), p.621 
\bibitem{}Draine, B. T., \& Lazarian, A. 1998, ApJ, 508, 157 (DL98)
\bibitem{dummy}Draine, B. T., Roberge, W. G. \& Dalgarno, A. 1983, ApJ, 264, 485 
\bibitem{dummy}Draine, B.T., \& Salpeter, E.E. 1979, ApJ, 231, 77 
\bibitem{dummy}Draine, B.T., \& Weingartner, J.C. 1996, ApJ 470, 551 

\bibitem{key-14}Ellison, D.C., Drury, L. O'C. \& Meyer, J.-P. 1997, ApJ, 487, 197 
\bibitem{}Elmegreen, \& Fiebig 1993, A\&A, 270, 397
\bibitem{dummy}Epstein, R. I. 1980, MNRAS, 193, 723
\bibitem{key-22}Foote, E. A. \& Kulsrud, R. M. 1979, ApJ, 233, 302
\bibitem{dummy}Gold, T. 1952, MNRAS, 112, 215
\bibitem{Goldreich Sridhar 1995}Goldreich, P. \& Sridhar, S. 1995, ApJ, 438, 763 
 \bibitem{dummy}Greenberg, J. M. \& Yencha, A. J. 1973, in IAU Symp 52, p. 369
\bibitem{key-37}Higdon, J. C. 1984, ApJ, 285, 109
\bibitem{}Jokipii, J. R. 1966, ApJ, 146, 480 
\bibitem{dummy}Jones, A.P., Tielens, A.G.G.M. \& Hollenbach, D.J. 1996, ApJ, 469,
740 
\bibitem{dummy}Kim, S. H., Martin, P. G., \& Hendry, P. D. 1994, ApJ, 422, 164 
\bibitem{}Kulsrud, R. M., \& Pearce, W. P., 1969, ApJ, 156, 445 
\bibitem{dummy}Kusaka, T., Nakano, T., \& Hayashi, C., 1970, Prog. Theor. Phys.,
44, 1580 
\bibitem{dummy}Hildebrand, R. H., Davidson, J. A., Dotson, J. L., Dowell, C. D.,
Novak, G. \& Vailancourt, J. E. 2000, ASP, 112, 1215
\bibitem{dummy}Lazarian, A. 1994, MNRAS, 268, 713 
\bibitem{dummy}Lazarian, A. 1995a, ApJ, 453, 229 
\bibitem{dummy}Lazarian, A. 1995b, MNRAS, 274, 679 
\bibitem{dummy}Lazarian, A. 1996, in ASP Conf. Proc. 97, Polarimetry of the Interstellar
Medium, eds. W. G. Roberge \& D. C. B. Whittet (San Francisco: ASP),
425 
\bibitem{dummy}Lazarian, A. 1997, ApJ, 483, 296 
\bibitem{key-27}Lazarian, A. 1999, Plasma Turbulence and Energetic Particles in Astrophysics,
Proceedings of the International Conference, Eds.: Michal Ostrowski,
Reinhard Schlickeiser, p. 28
\bibitem{dummy}Lazarian, A. 2000, in {}``Cosmic Evolution and Galaxy Formation'',
ASP, v.215, eds. Jose Franco, Elena Terlevich, Omar Lopez-Cruz, Itziar
Aretxaga, p. 69 
\bibitem{key-19}Lazarian, A. 2003, Journal of Quantitative Spectroscopy and Radiative
Transfer, 79, 881
\bibitem{dummy}Lazarian, A. \& Efroimsky, M. 1996, ApJ, 466, 274 
\bibitem{dummy}Lazarian, A. \& Efroimsky, M. 1999, MNRAS, 303, 673L 
\bibitem{dummy}Lazarian, A. \& Draine, B.T. 1997, ApJ, 487, 248 
\bibitem{dummy}Lazarian, A. \& Draine, B.T. 1999a, ApJ, 516, L37 
\bibitem{dummy}Lazarian, A. \& Draine, B.T. 1999b, ApJ, 520, L67 
\bibitem{dummy}Lazarian, A. \& Pogosyan, D. 2000, ApJ, 537, 720 
\bibitem{dummy}Lazarian, A., Vishniac, E., 1999, ApJ, 517, 700 
\bibitem{} Lazarian, A., Vishniac, E. \& Cho, J. 2004, ApJ, 603, 180
\bibitem{key-20}Lazarian, A. \& Yan, H., 2002a, ApJ, 566, 105L (LY02)
\bibitem{key-31}Lazarian, A. \& Yan, H., 2002b, Best of Science, astro-ph/0205283
\bibitem{}Lazarian, A. \& Yan, H. 2003, to appear in the proceedings of Astrophysics Dust, astro-ph/0311370
\bibitem{dummy}Lepp, S. 1992, in Astrochemistry of Cosmic Phenomena,
IAU Symp. 150, ed. P. Singh (Dordrecht: Kluwer), p. 471
\bibitem{dummy}Lithwick, Y. \& Goldreich, P. 2001, ApJ, 562, 279 
\bibitem{dummy}Maron, J. \& Goldreich, P. 2001, ApJ, 554, 1175 
\bibitem{dummy}McCall, B.J., Huneycutt, A.J., Saykally, R.J.,
Geballe, T.R., Djuric, N., Dunn, G.H., et al.\ 2003, 
Nature, 422, 500
\bibitem{key-15}Melrose, D.B. 1980, Plasma Astrophysics (New York: Gordon \& Breach). 
\bibitem{}Montgomery, D. \& Matthaeus, W. 1981, Phys. Fluids, 24 825
\bibitem{dummy}Mukherjee, P., Jones, A.W., Kneissl, R., Lasenby, A.N. 2001, MNRAS, 320, 224
\bibitem{}Nishi, R., Nakano, T., \& Umebayashi, T. 1991, ApJ, 368, 181
\bibitem{dummy}Ossenkopf, V. 1993, A\&A 280, 617 
\bibitem{key-16}Pryadko, J.M. \& Petrosian, V. 1997, ApJ, 482, 774
\bibitem{dummy}Purcell, E.M. 1969, Physica 41, 100 
\bibitem{key-17}Purcell, E.M. 1979, ApJ, 231, 404 
\bibitem{dummy}Ruffle, D.P., Hartquist, T.W., Rawlings, J.M.C.,
\& Williams, D.A. 1998, A\&A, 334, 678
\bibitem{key-36}Scalo, J. M. 1987, In Interstellar processes, Proceedings of the
Symposium, eds. D.J. Hollenback and Harley A. Thronson, Jr., p. 349 
\bibitem{dummy}Schlickeiser, R. \& Achatz, U. 1993, J. Plasma Phy. 49, 63 
\bibitem{dummy}Schlickeiser, R. \& Miller, J. A. 1998, ApJ, 492, 352 
\bibitem{key-35}Schlickeiser, R. 2002, \emph{Cosmic Ray Astrophysics} (Springer-Verlag Berlin Heidelberg)
\bibitem{dummy}Schutte, W. A. \& Greenberg, J. M. 1991, A \& A 244, 190
\bibitem{key-39}Shebalin, J. V., Matthaeus, W. H., \& Montgomery, D. 1983, J. Plasma Phys., 29, 525
\bibitem{dummy}Spitzer, L. \& McGlynn, T.A. 1979, ApJ, 231, 417 
\bibitem{dummy}Spitzer, L., 1978, Physical Processes in the Interstellar Medium (New York:Wiley) 
\bibitem{dummy}Stanimirovic, S. \& Lazarian, A. 2001, ApJ, 551, L53 
\bibitem{dummy}V\"olk, H.J., Jones, F.C., Morfill, G.E., \& Roser, S. 1980
A\&A, 85, 316 
\bibitem{key-23}Wakker, B. P. \& Mathis, J. S. 2000, ApJ, 544, 107L
\bibitem{dummy}Weidenschilling, S.J. \& Ruzmaikina, T.V. 1994, ApJ, 430, 713 
\bibitem{dummy}Weingartner, J.C. \& Draine, B.T. 2001a, ApJ, 553, 581  (WD01a)
\bibitem{dummy}Weingartner, J.C. \& Draine, B.T. 2001b, ApJS, 134, 263 (WD01b)
\bibitem{dummy}Weingartner, J.C., \& Draine, B.T. 2002, ApJ, 563, 842
\bibitem{dummy}Welty, D.E., Hobbs, L.M., Lauroesch, J.T., Morton, D.C.,
  Spitzer, L., \& York, D.G. 1999, ApJS, 124, 465
\bibitem{key-13}Yan, H. \& Lazarian, A. 2002, Phy. Rev. Lett, 89, 281102 (YL02)
\bibitem{key-32}Yan, H. \& Lazarian, A. 2003, ApJ, 592, 33L (YL03)
\bibitem{}Yan, H. \& Lazarian, A. 2004, accepted to ApJ
\bibitem{}Zank, G.P. \& Mattaeus, W.H. 1992, J. Plasma Phys. 48, 85
\end{thebibliography}
\end{document}